\documentclass[twocolumn,times]{aastex631}
\usepackage[utf8]{inputenc}
\usepackage[nolist]{acronym}
\usepackage{listings}

\providecommand{\acrolowercase}[1]{\lowercase{#1}}

\begin{acronym}
\acro{2D}[2D]{two\nobreakdashes-dimensional}
\acro{2+1D}[2+1D]{2+1\nobreakdashes--dimensional}
\acro{2MRS}[2MRS]{2MASS Redshift Survey}
\acro{3D}[3D]{three\nobreakdashes-dimensional}
\acro{2MASS}[2MASS]{Two Micron All Sky Survey}
\acro{AdVirgo}[AdVirgo]{Advanced Virgo}
\acro{AMI}[AMI]{Arcminute Microkelvin Imager}
\acro{AGN}[AGN]{active galactic nucleus}
\acroplural{AGN}[AGN\acrolowercase{s}]{active galactic nuclei}
\acro{aLIGO}[aLIGO]{Advanced \acs{LIGO}}
\acro{ASKAP}[ASKAP]{Australian \acl{SKA} Pathfinder}
\acro{ATCA}[ATCA]{Australia Telescope Compact Array}
\acro{ATLAS}[ATLAS]{Asteroid Terrestrial-impact Last Alert System}
\acro{AWS}[AWS]{Amazon Web Services}
\acro{BAT}[BAT]{Burst Alert Telescope\acroextra{ (instrument on \emph{Swift})}}
\acro{BATSE}[BATSE]{Burst and Transient Source Experiment\acroextra{ (instrument on \acs{CGRO})}}
\acro{BAYESTAR}[BAYESTAR]{BAYESian TriAngulation and Rapid localization}
\acro{BBH}[BBH]{binary black hole}
\acro{BHBH}[BHBH]{\acl{BH}\nobreakdashes--\acl{BH}}
\acro{BH}[BH]{black hole}
\acro{BNS}[BNS]{binary neutron star}
\acro{CARMA}[CARMA]{Combined Array for Research in Millimeter\nobreakdashes-wave Astronomy}
\acro{CASA}[CASA]{Common Astronomy Software Applications}
\acro{CBCG}[CBCG]{Compact Binary Coalescence Galaxy}
\acro{CFH12k}[CFH12k]{Canada--France--Hawaii $12\,288 \times 8\,192$ pixel CCD mosaic\acroextra{ (instrument formerly on the Canada--France--Hawaii Telescope, now on the \ac{P48})}}
\acro{CLU}[CLU]{Census of the Local Universe}
\acro{CRTS}[CRTS]{Catalina Real-time Transient Survey}
\acro{CTIO}[CTIO]{Cerro Tololo Inter-American Observatory}
\acro{CBC}[CBC]{compact binary coalescence}
\acro{CCD}[CCD]{charge coupled device}
\acro{CDF}[CDF]{cumulative distribution function}
\acro{CDS}[CDS]{Centre de Données astronomiques de Strasbourg}
\acro{CGRO}[CGRO]{Compton Gamma Ray Observatory}
\acro{CMB}[CMB]{cosmic microwave background}
\acro{CRLB}[CRLB]{Cram\'{e}r\nobreakdashes--Rao lower bound}
\acro{cWB}[\acrolowercase{c}WB]{Coherent WaveBurst}
\acro{DASWG}[DASWG]{Data Analysis Software Working Group}
\acro{DBSP}[DBSP]{Double Spectrograph\acroextra{ (instrument on \acs{P200})}}
\acro{DCT}[DCT]{Discovery Channel Telescope}
\acro{DECAM}[DECam]{Dark Energy Camera\acroextra{ (instrument on the Blanco 4\nobreakdashes-m telescope at \acs{CTIO})}}
\acro{DES}[DES]{Dark Energy Survey}
\acro{DFT}[DFT]{discrete Fourier transform}
\acro{EC2}[EC2]{Elastic Compute Cloud}
\acro{EM}[EM]{electromagnetic}
\acro{ER8}[ER8]{eighth engineering run}
\acro{FD}[FD]{frequency domain}
\acro{FAR}[FAR]{false alarm rate}
\acro{FFT}[FFT]{fast Fourier transform}
\acro{FIR}[FIR]{finite impulse response}
\acro{FITS}[FITS]{Flexible Image Transport System}
\acro{F2}[F2]{FLAMINGOS\nobreakdashes-2}
\acro{FLOPS}[FLOPS]{floating point operations per second}
\acro{FOV}[FOV]{field of view}
\acroplural{FOV}[FOV\acrolowercase{s}]{fields of view}
\acro{FTN}[FTN]{Faulkes Telescope North}
\acro{FWHM}[FWHM]{full width at half-maximum}
\acro{GBM}[GBM]{Gamma-ray Burst Monitor\acroextra{ (instrument on \emph{Fermi})}}
\acro{GCN}[GCN]{Gamma-ray Coordinates Network}
\acro{GiST}[GiST]{Generalized Search Tree}

\acro{GLADE}[GLADE]{Galaxy List for the Advanced Detector Era}
\acro{GMOS}[GMOS]{Gemini Multi-object Spectrograph\acroextra{ (instrument on the Gemini telescopes)}}
\acro{GRB}[GRB]{gamma-ray burst}
\acro{GraceDB}[GraceDB]{GRAvitational-wave Candidate Event DataBase}
\acro{GROWTH}[GROWTH]{Global Relay of Observatories Watching Transients Happen}
\acro{GSC}[GSC]{Gas Slit Camera}
\acro{GSL}[GSL]{GNU Scientific Library}
\acro{GTC}[GTC]{Gran Telescopio Canarias}
\acro{GW}[GW]{gravitational wave}
\acro{GWGC}[GWGC]{Gravitational Wave Galaxy Catalogue}
\acro{HAWC}[HAWC]{High\nobreakdashes-Altitude Water \v{C}erenkov Gamma\nobreakdashes-Ray Observatory}
\acro{HCT}[HCT]{Himalayan Chandra Telescope}
\acro{HEALPix}[HEALP\acrolowercase{ix}]{Hierarchical Equal Area isoLatitude Pixelization}
\acro{HEASARC}[HEASARC]{High Energy Astrophysics Science Archive Research Center}
\acro{HETE}[HETE]{High Energy Transient Explorer}
\acro{HFOSC}[HFOSC]{Himalaya Faint Object Spectrograph and Camera\acroextra{ (instrument on \acs{HCT})}}
\acro{HiPS}[HiPS]{hierarchical progressive surveys}
\acro{HMXB}[HMXB]{high\nobreakdashes-mass X\nobreakdashes-ray binary}
\acroplural{HMXB}[HMXB\acrolowercase{s}]{high\nobreakdashes-mass X\nobreakdashes-ray binaries}
\acro{HSC}[HSC]{Hyper Suprime\nobreakdashes-Cam\acroextra{ (instrument on the 8.2\nobreakdashes-m Subaru telescope)}}
\acro{HTM}[HTM]{Hierarchical Triangular Mesh}
\acro{IACT}[IACT]{imaging atmospheric \v{C}erenkov telescope}
\acro{ICRS}[ICRS]{International Celestial Reference System}
\acro{IIR}[IIR]{infinite impulse response}
\acro{IMACS}[IMACS]{Inamori-Magellan Areal Camera \& Spectrograph\acroextra{ (instrument on the Magellan Baade telescope)}}
\acro{IMR}[IMR]{inspiral-merger-ringdown}
\acro{IPAC}[IPAC]{Infrared Processing and Analysis Center}
\acro{IPN}[IPN]{InterPlanetary Network}
\acro{IPTF}[\acrolowercase{i}PTF]{intermediate \acl{PTF}}
\acro{IRAC}[IRAC]{Infrared Array Camera}
\acro{ISM}[ISM]{interstellar medium}
\acro{ISS}[ISS]{International Space Station}
\acro{IVOA}[IVOA]{International Virtual Observatory Alliance}
\acro{KAGRA}[KAGRA]{KAmioka GRAvitational\nobreakdashes-wave observatory}
\acro{KDE}[KDE]{kernel density estimator}
\acro{KN}[KN]{kilonova}
\acroplural{KN}[KNe]{kilonovae}
\acro{LAT}[LAT]{Large Area Telescope}
\acro{LCO}[LCO]{Las Cumbres Observatory}
\acro{LCOGT}[LCOGT]{Las Cumbres Observatory Global Telescope}
\acro{LHO}[LHO]{\ac{LIGO} Hanford Observatory}
\acro{LIB}[LIB]{LALInference Burst}
\acro{LIGO}[LIGO]{Laser Interferometer \acs{GW} Observatory}
\acro{llGRB}[\acrolowercase{ll}GRB]{low\nobreakdashes-luminosity \ac{GRB}}
\acro{LLOID}[LLOID]{Low Latency Online Inspiral Detection}
\acro{LLO}[LLO]{\ac{LIGO} Livingston Observatory}
\acro{LMI}[LMI]{Large Monolithic Imager\acroextra{ (instrument on \ac{DCT})}}
\acro{LOFAR}[LOFAR]{Low Frequency Array}
\acro{LOS}[LOS]{line of sight}
\acroplural{LOS}[LOSs]{lines of sight}
\acro{LMC}[LMC]{Large Magellanic Cloud}
\acro{LSB}[LSB]{long, soft burst}
\acro{LSC}[LSC]{\acs{LIGO} Scientific Collaboration}
\acro{LSO}[LSO]{last stable orbit}
\acro{LSST}[LSST]{Large Synoptic Survey Telescope}
\acro{LT}[LT]{Liverpool Telescope}
\acro{LTI}[LTI]{linear time invariant}
\acro{MAP}[MAP]{maximum a posteriori}
\acro{MBTA}[MBTA]{Multi-Band Template Analysis}
\acro{MCMC}[MCMC]{Markov chain Monte Carlo}
\acro{MLE}[MLE]{\ac{ML} estimator}
\acro{ML}[ML]{maximum likelihood}
\acro{MMA}[MMA]{multi-messenger astronomy}
\acro{MOC}[MOC]{multi-order coverage map}
\acro{MOU}[MOU]{memorandum of understanding}
\acroplural{MOU}[MOUs]{memoranda of understanding}
\acro{MWA}[MWA]{Murchison Widefield Array}
\acro{NED}[NED]{NASA/IPAC Extragalactic Database}
\acro{NIR}[NIR]{near infrared}
\acro{NSBH}[NSBH]{neutron star\nobreakdashes--black hole}
\acro{NSBH}[NSBH]{\acl{NS}\nobreakdashes--\acl{BH}}
\acro{NSF}[NSF]{National Science Foundation}
\acro{NSNS}[NSNS]{\acl{NS}\nobreakdashes--\acl{NS}}
\acro{NS}[NS]{neutron star}
\acro{O1}[O1]{\acl{aLIGO}'s first observing run}
\acro{O2}[O2]{\acl{aLIGO}'s and \acl{AdVirgo}'s second observing run}
\acro{O3}[O3]{\acl{aLIGO}'s and \acl{AdVirgo}'s third observing run}
\acro{oLIB}[\acrolowercase{o}LIB]{Omicron+\acl{LIB}}
\acro{ORM}[ORM]{object relational mapper}
\acro{OT}[OT]{optical transient}
\acro{P48}[P48]{Palomar 48~inch Oschin telescope}
\acro{P60}[P60]{robotic Palomar 60~inch telescope}
\acro{P200}[P200]{Palomar 200~inch Hale telescope}
\acro{PC}[PC]{photon counting}
\acro{PESSTO}[PESSTO]{Public ESO Spectroscopic Survey of Transient Objects}
\acro{PSD}[PSD]{power spectral density}
\acro{PSF}[PSF]{point-spread function}
\acro{PS1}[PS1]{Pan\nobreakdashes-STARRS~1}
\acro{PTF}[PTF]{Palomar Transient Factory}
\acro{QUEST}[QUEST]{Quasar Equatorial Survey Team}
\acro{RAPTOR}[RAPTOR]{Rapid Telescopes for Optical Response}
\acro{RDS}[RDS]{Relational Database Service}
\acro{REU}[REU]{Research Experiences for Undergraduates}
\acro{RMS}[RMS]{root mean square}
\acro{ROTSE}[ROTSE]{Robotic Optical Transient Search}
\acro{Rubin}[Rubin]{Vera C. Rubin Observatory}
\acro{S5}[S5]{\ac{LIGO}'s fifth science run}
\acro{S6}[S6]{\ac{LIGO}'s sixth science run}
\acro{SAA}[SAA]{South Atlantic Anomaly}
\acro{SHB}[SHB]{short, hard burst}
\acro{SHGRB}[SHGRB]{short, hard \acl{GRB}}
\acro{SKA}[SKA]{Square Kilometer Array}
\acro{SMT}[SMT]{Slewing Mirror Telescope\acroextra{ (instrument on \acs{UFFO} Pathfinder)}}
\acro{SNR}[S/N]{signal\nobreakdashes-to\nobreakdashes-noise ratio}
\acro{SSC}[SSC]{synchrotron self\nobreakdashes-Compton}
\acro{SDSS}[SDSS]{Sloan Digital Sky Survey}
\acro{SED}[SED]{spectral energy distribution}
\acro{SFR}[SFR]{star formation rate}
\acro{SGRB}[SGRB]{short \acl{GRB}}
\acro{SN}[SN]{supernova}
\acroplural{SN}[SN\acrolowercase{e}]{supernova}
\acro{SNIa}[\acs{SN}\,I\acrolowercase{a}]{Type~Ia \ac{SN}}
\acroplural{SNIa}[\acsp{SN}\,I\acrolowercase{a}]{Type~Ic \acp{SN}}
\acro{SNIcBL}[\acs{SN}\,I\acrolowercase{c}\nobreakdashes-BL]{broad\nobreakdashes-line Type~Ic \ac{SN}}
\acroplural{SNIcBL}[\acsp{SN}\,I\acrolowercase{c}\nobreakdashes-BL]{broad\nobreakdashes-line Type~Ic \acp{SN}}
\acro{SP-GiST}[SP-GiST]{space-partitioned GiST}
\acro{SQL}[SQL]{Structured Query Language}
\acro{SVD}[SVD]{singular value decomposition}
\acro{TAROT}[TAROT]{T\'{e}lescopes \`{a} Action Rapide pour les Objets Transitoires}
\acro{TDOA}[TDOA]{time delay on arrival}
\acroplural{TDOA}[TDOA\acrolowercase{s}]{time delays on arrival}
\acro{TD}[TD]{time domain}
\acro{TOA}[TOA]{time of arrival}
\acroplural{TOA}[TOA\acrolowercase{s}]{times of arrival}
\acro{TOM}[TOM]{target observation manager}
\acro{ToO}[ToO]{target of opportunity}
\acroplural{TOO}[TOO\acrolowercase{s}]{targets of opportunity}
\acro{UFFO}[UFFO]{Ultra Fast Flash Observatory}
\acro{UHE}[UHE]{ultra high energy}
\acro{UMN}[UMN]{University of Minnesota}
\acro{UVOT}[UVOT]{UV/Optical Telescope\acroextra{ (instrument on \emph{Swift})}}
\acro{VHE}[VHE]{very high energy}
\acro{VISTA}[VISTA@ESO]{Visible and Infrared Survey Telescope}
\acro{VLA}[VLA]{Karl G. Jansky Very Large Array}
\acro{VLT}[VLT]{Very Large Telescope}
\acro{VO}[VO]{Virtual Observatory}
\acro{VST}[VST@ESO]{\acs{VLT} Survey Telescope}
\acro{WAM}[WAM]{Wide\nobreakdashes-band All\nobreakdashes-sky Monitor\acroextra{ (instrument on \emph{Suzaku})}}
\acro{WCS}[WCS]{World Coordinate System}
\acro{WSS}[w.s.s.]{wide\nobreakdashes-sense stationary}
\acro{XRF}[XRF]{X\nobreakdashes-ray flash}
\acroplural{XRF}[XRF\acrolowercase{s}]{X\nobreakdashes-ray flashes}
\acro{XRT}[XRT]{X\nobreakdashes-ray Telescope\acroextra{ (instrument on \emph{Swift})}}
\acro{ZTF}[ZTF]{Zwicky Transient Facility}
\end{acronym}

\received{2021 December 13}
\accepted{2022 March 1}
\published{2022 April 12}

\shorttitle{HEALPix Alchemy}
\shortauthors{Singer, Parazin, Coughlin, et al.}

\begin{document}

\title{HEALPix Alchemy: Fast All-Sky Geometry and Image Arithmetic in a Relational Database for Multimessenger Astronomy Brokers}

\correspondingauthor{Leo P. Singer}
\email{leo.p.singer@nasa.gov}

\author[0000-0001-9898-5597]{Leo P. Singer}
\affiliation{Astroparticle Physics Laboratory, NASA Goddard Space Flight Center, Code 661, Greenbelt, MD 20771, USA}

\author[0000-0002-3155-0385]{B. Parazin}
\affiliation{Northeastern University, Boston, MA 02115, USA}
\affiliation{School of Physics and Astronomy, University of Minnesota, Minneapolis, MN 55455, USA}

\author[0000-0002-8262-2924]{Michael W. Coughlin}
\affiliation{School of Physics and Astronomy, University of Minnesota, Minneapolis, MN 55455, USA}

\author[0000-0002-7777-216X]{Joshua S. Bloom}
\affiliation{Department of Astronomy, University of California,
Berkeley, CA 94720, USA}
\affiliation{Lawrence Berkeley National Laboratory, 1 Cyclotron Road,
MS 50B-4206, Berkeley, CA 94720, USA}

\author[0000-0002-7183-0410]{Arien Crellin-Quick}
\affiliation{Department of Astronomy, University of California,
Berkeley, CA 94720, USA}

\author[0000-0003-3461-8661]{Daniel A. Goldstein}
\affiliation{Weights and Biases, Inc., 1479 Folsom Street, San Francisco, CA 90063, USA}

\author[0000-0001-9276-1891]{Stéfan van der Walt}
\affiliation{Berkeley Institute for Data Science, University of California Berkeley, Berkeley, CA 94720, USA}

\begin{abstract}
Efficient searches for electromagnetic counterparts to gravitational wave, high-energy neutrino, and gamma-ray burst events demand rapid processing of image arithmetic and geometry set operations in a database to cross-match galaxy catalogs, observation footprints, and all-sky images. Here we introduce HEALPix Alchemy, an open-source, pure Python implementation of a set of methods that enables rapid all-sky geometry calculations. HEALPix Alchemy is built upon HEALPix, a spatial indexing strategy that is widely used in astronomical databases as well as the native format of LIGO-Virgo-KAGRA gravitational-wave sky localization maps. Our approach leverages new multirange types built into the PostgreSQL 14 database engine. This enables fast all-sky queries against probabilistic multimessenger event localizations and telescope survey footprints. Questions such as ``What are the galaxies contained within the 90\% credible region of an event?'' and ``What is the rank-ordered list of the fields within an observing footprint with the highest probability of containing the event?'' can be performed in less than a few seconds on commodity hardware using off-the-shelf cloud-managed database implementations without server-side database extensions. Common queries scale roughly linearly with the number of telescope pointings. As the number of fields grows into the hundreds or thousands, HEALPix Alchemy is orders of magnitude faster than other implementations. HEALPix Alchemy is now used as the spatial geometry engine within SkyPortal, which forms the basis of the Zwicky Transient Facility transient marshal, called Fritz.
\end{abstract}


\section{Introduction}

The multimessenger view of astrophysical transients expands our understanding of the origin and nature of compact objects, relativistic outflows, and nucleosynthesis. However, the discovery and study of \ac{EM} counterparts associated with \ac{GW} events, high-energy neutrino sources, and \acp{GRB} has proven challenging given the wide localizations of such events relative to the narrow \acp{FOV} of optical and radio follow-up facilities: the position uncertainties for \ac{GW}, \ac{GRB}, and neutrino events are typically tens to thousands of square degrees \citep{2015ApJS..216...32C,2017APh....92...30A,2018LRR....21....3A,2022ApJ...924...54P}, whereas the \acp{FOV} of optical telescopes that are sensitive to the \ac{EM} counterparts are rarely more than tens of square degrees. 

Thankfully, innovations in automation, scheduling, and coordination have made it feasible to observe and then reobserve wide swaths of the sky to search for variability. Survey telescopes such as the \acl{ZTF} (\acsu{ZTF}; \citealt{2019PASP..131a8002B,2019PASP..131g8001G,2019PASP..131a8003M,2020PASP..132c8001D}) and the upcoming \aclu{Rubin} \citep{2019ApJ...873..111I} provide wide-field, deep, semiautonomous rapidly slewing observing capabilities. Such facilities have been able to cover entire localization regions in hours to days following an event \citep[e.g.,][]{2020ApJ...905..145K}. Smaller \ac{FOV} facilities have been used to selectively target nearby galaxies in event localization regions. Alert brokers have been developed to ingest transient detection alerts in real time in order to filter, collate, and present alerts according to user-programmable filtering rules \citep[e.g.,][]{2019A&A...631A.147N,2019RNAAS...3...26S,2021AJ....161..242F,2021AJ....161..107M,2021MNRAS.501.3272M,pitt-google-broker}. Scientists submit targets for further observations on other facilities using a \acl{TOM} (\acsu{TOM}; \citealt{tom-toolkit-workshop}). Marshal applications may be used to analyze, view, and collaborate on all of the observations related to one or a collection of objects under study \citep[e.g.,][]{2019PASP..131c8003K,2019JOSS....4.1247V}. Science working groups can select different targets of interest to feed to robotic follow-up networks like \acl{LCO} (\acsu{LCO}; \citealt{2013PASP..125.1031B}), queue-scheduled 8m-class observatories like Gemini, and future very large aperture telescopes.

While most of the software and hardware components are already in place to enable EM follow-up, large position uncertainties require multimessenger astronomy brokers and marshals to support special kinds of spatial queries that are not common for other science cases \citep[e.g.,][]{2020NatAs...4..550C,2020ApJ...894..127W}. Localizations of \ac{GW} events from the \acl{LIGO} (\acsu{LIGO}; \citealt{2015CQGra..32g4001L}), Virgo \citep{2015CQGra..32b4001A}, and the \acl{KAGRA} (\acsu{KAGRA}; \citealt{2021PTEP.2021eA101A}), and of \acp{GRB} from the Fermi \acl{GBM} (\acsu{GBM}; \citealt{2009ApJ...702..791M,2020ApJ...895...40G}) take the form of all-sky probability map images. To plan a tiled \ac{ToO} search for the \ac{EM} counterpart, brokers need to be able to rapidly calculate the probability contained within the footprint of an observation or the union of many tiled observations. To rank potential candidates, marshals need to be able to cross-match catalogs of point sources with the probability maps. In short, multimessenger applications require cross-matches of points, regions, and images.

\subsection{HEALPix}

Several technologies are widely used to accelerate geometry processing of points and regions in astronomical information systems. The \acl{HEALPix} (\acsu{HEALPix}; \citealt{2005ApJ...622..759G}) has been especially influential in this area. \ac{HEALPix} is an all-sky projection and spatial indexing method that was originally designed for \ac{CMB} analysis, where uniform sky sampling without artifacts at projection boundaries is essential. One of the authors of this work \citep{2016PhRvD..93b4013S} later introduced \ac{HEALPix} to the \ac{GW} community as the standard format for \ac{GW} localizations for similar reasons.

At any given resolution, \ac{HEALPix} tessellates the unit sphere and addresses each tile with an integer pixel index. \ac{HEALPix} subdivides the unit sphere into a multi-resolution tree of nested pixels, much like a quadtree subdivides a bounded region of a 2D plane or an octree subdivides a bounded region of 3D space in classic computer graphics and numerical astrophysics applications. The carefully designed algebraic properties of HEALPix indices are such that tiles that are siblings in the HEALPix tree have adjacent pixel indices. Because spatially neighboring tiles tend to have neighboring addresses, \ac{HEALPix} indices are readily useful as database indices to speed up spatial queries.

\begin{figure}
    \includegraphics[width=\columnwidth]{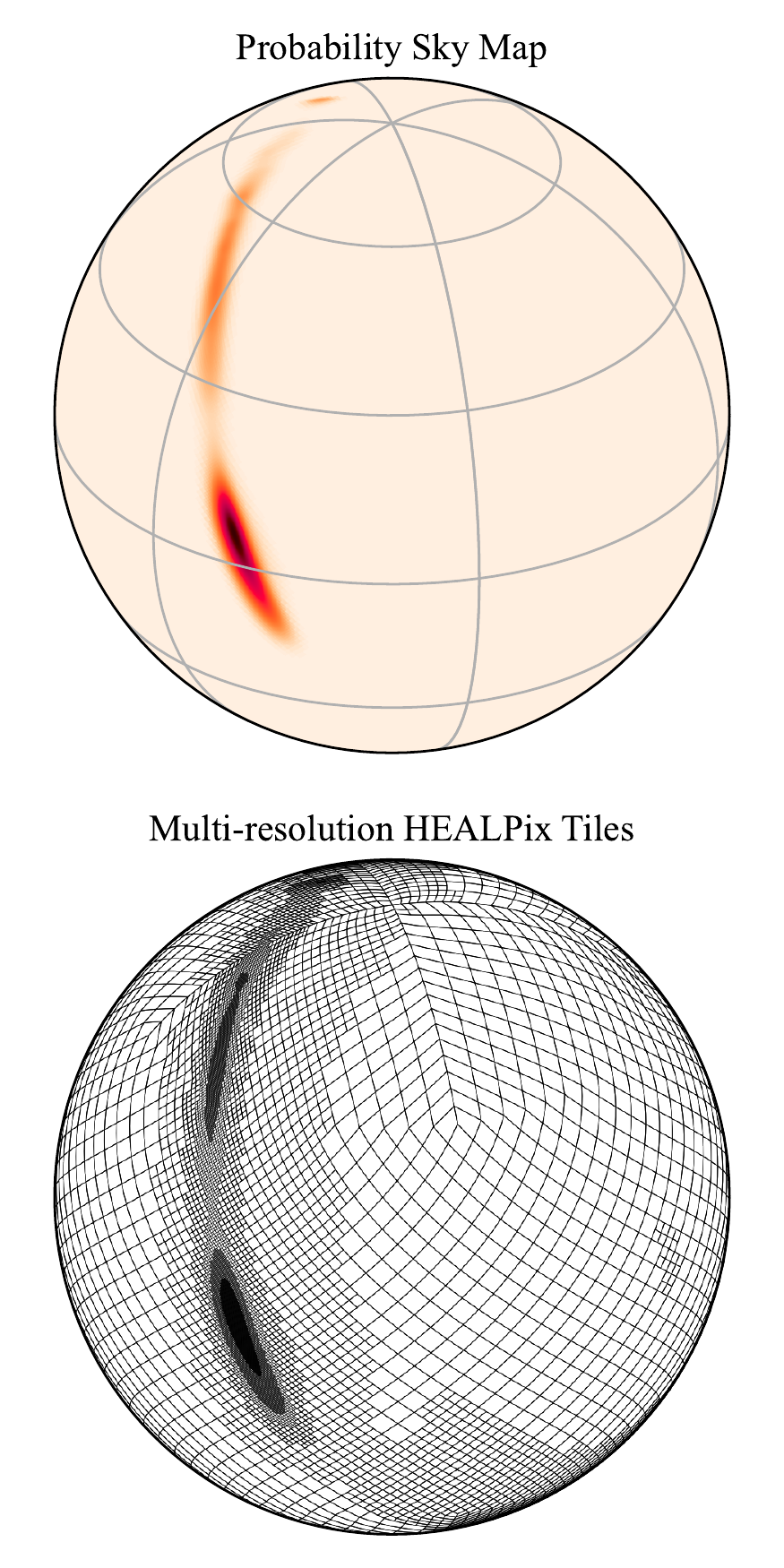}
    \caption{An example of the multi-resolution \ac{HEALPix} sampling scheme that is used in \ac{GW} localizations. The top panel shows the localization of LIGO/Virgo event GW200115\_042309 (\citealt{2021ApJ...915L...5A}; GraceDB ID S200115j) as a heat map image; darker, deeper colors represent higher probability density. The bottom panel shows the boundaries of the multi-resolution HEALPix tiles on which the sky map was sampled.}
    \label{fig:skymap-mesh}
\end{figure}

These properties \citep{2015A&A...580A.132R} are central to how \ac{GW} probability maps are produced and stored. \ac{GW} sky maps are sampled on an adaptively refined \ac{HEALPix} grid, with pixel density roughly proportional to probability density (\citealt{2016PhRvD..93b4013S}; see Fig~\ref{fig:skymap-mesh} for an example). This saves a substantial amount of time when generating \ac{GW} sky maps and a significant amount of bandwidth and storage when broadcasting the localizations to astronomers.

The serialization format for \ac{GW} sky maps is based upon \aclp{MOC} (\acsp{MOC}\acused{MOC}; \citealt{2014ivoa.spec.0602F,2019ivoa.spec.1007F}), an \ac{IVOA} specification for encoding footprints of observations or surveys as multi-resolution HEALPix bit masks to enable fast spatial unions and intersections. \acp{MOC} are used extensively in the Aladin sky atlas \citep{2000A&AS..143...33B,2014ASPC..485..277B} and many other \ac{VO} tools and platforms. \citet{Greco:20191+} brought \acp{MOC} to prominence in the \ac{GW} community by adding \ac{MOC} contouring of \ac{GW} probability maps and cross-matching with catalogs to Aladin. The hierarchical nature of \ac{HEALPix} also underlies the \ac{IVOA} \ac{HiPS} standard \citep{2015A&A...578A.114F,2017ivoa.spec.0519F}, an astronomy map tile technology that enables interactive panning and zooming, similar to Google Maps, in Aladin.

\subsection{HEALPix and Spatial Indices in Databases}

There are a multitude of software packages that add \ac{HEALPix} or similar spatial indices to common relational databases, including H3C \citep{2013ASPC..475..227L}, pg\_healpix \citep{pghealpix}, Q3C \citep{2006ASPC..351..735K,2019ascl.soft05008K}, \acl{HTM} (\acsu{HTM}; \citealt{2007cs........1164S}), and pgSphere \citep{2004ASPC..314..225C}. There is significant technological overlap with geospatial packages like PostGIS \citep{obe2021postgis}. However, with the exception of PostGIS, these extensions and technologies do not naturally handle the image queries and arithmetic needed for directly processing multimessenger localizations.

Furthermore, all of these software packages are binary database extensions that must be specially installed or enabled on the server, making them difficult to deploy on robust, fault-tolerant, fully managed database services in the cloud, like Amazon \ac{RDS} (\url{https://aws.amazon.com/rds/}), Google Cloud SQL (\url{https://cloud.google.com/sql}), and Microsoft Azure Database for PostgreSQL (\url{https://azure.microsoft.com/services/postgresql/}). Out of all of the extensions listed above, only PostGIS is supported by these managed database services.

\subsection{HEALPix in Python}

There are many high-quality Python implementations of HEALPix. We list a few relevant ones here.

Healpy \citep{2019JOSS....4.1298Z} wraps the official HEALPix \citep{2005ApJ...622..759G} C++ library as a NumPy \citep{2020Natur.585..357H} C extension. It is available as a stand-alone Python package from the Python Package Index, but is also included in the official polyglot HEALPix bundle. Healpy is the tool of choice for \ac{CMB} analysis in Python because it exposes the underlying C++ library's capability to transform HEALPix data sets to and from the space of spherical harmonics.

The astropy-healpix \citep{2020ascl.soft11023R} project is a BSD-licensed Astropy-coordinated package with a high-level object-oriented interface and excellent integration with Astropy coordinates and units \citep{2013A&A...558A..33A,2018AJ....156..123A}. Notably, astropy-healpix is used by the Astropy-coordinated reproject package to provide high-quality image reprojection between \ac{HEALPix} and \aclp{WCS} (\acused{WCS}\acsp{WCS}; \citealt{2002A&A...395.1077C,2002A&A...395.1061G}). Behind the scenes, astropy-healpix wraps a C implementation of HEALPix adapted from Astrometry.net \citep{2010AJ....139.1782L}.

MOCPy \citep{2019ASPC..521..487B}, developed at the \ac{CDS}, is an Astropy-affiliated package that provides fast manipulation of \acp{MOC} in Python, also with a high-level object-oriented interface. Its HEALPix support comes from the cdshealpix Python package, which wraps \ac{CDS}'s implementation of HEALPix in the Rust programming language.

The most recent addition is mhealpy \citep{2021arXiv211111240M}, which combines some of the best features of Healpy, astropy-healpix, and MOCPy. The mhealpy package provides a unified object-oriented interface for conventional fixed-resolution HEALPix data sets (like Healpy and astropy-healpix) and multi-resolution data sets (like MOCPy). For \acp{MOC}, mhealpy supports not only region operations but also multi-resolution image arithmetic with a variety of options for normalization, making it a great choice for handling multi-resolution \ac{GW} and \ac{GRB} probability sky maps.

\subsection{HEALPix Alchemy}

One of several equivalent concrete data structures that can be used to encode multi-resolution \ac{HEALPix} geometry is an interval set or range set \citep{2015A&A...580A.132R} consisting of a set of disjoint ranges of integer pixel indices. In the range set representation, calculating the union or intersection of any number of spatial regions reduces to simply merging sorted lists of integers. The range set data structure cuts across many areas of data science, from \acp{GW} (the authors acknowledge Kipp Cannon's influential and elegant ligo-segments package, which has been one of the unsung heroes behind \ac{LIGO} and Virgo observational results for years; see \citealt{ligo-segments}) to bioinformatics \citep{10.1093/bioinformatics/btl647,10.1093/bioinformatics/btz615}, not to mention obvious applications in business software. Because of the multitude of applications (frankly, in fields that are better funded than astronomy), there is a wealth of software for fast, general-purpose processing of range sets. PostgreSQL 14 was recently released with a new built-in \emph{multirange} column type that maps perfectly onto the concept of \ac{HEALPix} range sets.

In this paper, we introduce HEALPix Alchemy, a pure Python package that extends the popular SQLAlchemy database toolkit for Python \citep{sqlalchemy} to add fast multi-resolution \ac{HEALPix} geometry on top of a PostgreSQL database using PostgreSQL 14 multiranges. HEALPix Alchemy accelerates queries involving cross-matches of points, regions, and images. Unlike traditional spatial indexing strategies, HEALPix Alchemy works with an unmodified PostgreSQL database service without any server-side extensions. HEALPix Alchemy can evaluate bulk queries involving unions of large numbers of regions ($\gtrsim$10) orders of magnitude faster than conventional, non-database multi-order HEALPix implementations like MOCPy. HEALPix Alchemy facilitates fast queries of \ac{GW} sky maps by directly exploiting their native multi-resolution sampling.

The organization of the paper is as follows. In Section~\ref{sec:healpix}, we review the fundamentals and algebraic properties of \ac{HEALPix}. In Section~\ref{sec:postgresql}, we summarize the new multirange support in PostgreSQL. In Section~\ref{sec:api}, we explain the design, interface, and usage of HEALPix Alchemy. In Section~\ref{sec:tutorial}, our sample code illustrates how to perform a variety of spatial queries that are important to a multimessenger broker or marshal application. Finally, in Section~\ref{sec:performance}, our benchmarks show that the HEALPix Alchemy approach is fast and scalable. The code is open source and publicly available on GitHub (\url{https://github.com/skyportal/healpix-alchemy}) and Zenodo \citep{healpix-alchemy-1.0.1}.

\section{HEALPix Fundamentals}
\label{sec:healpix}

We begin by summarizing \citet{2005ApJ...622..759G} to provide a brief overview of \ac{HEALPix}. \ac{HEALPix} is both an all-sky map projection and a spatial indexing method. \ac{HEALPix} divides and covers the unit sphere with equal-area tiles.

\begin{figure}
    \includegraphics[width=\columnwidth]{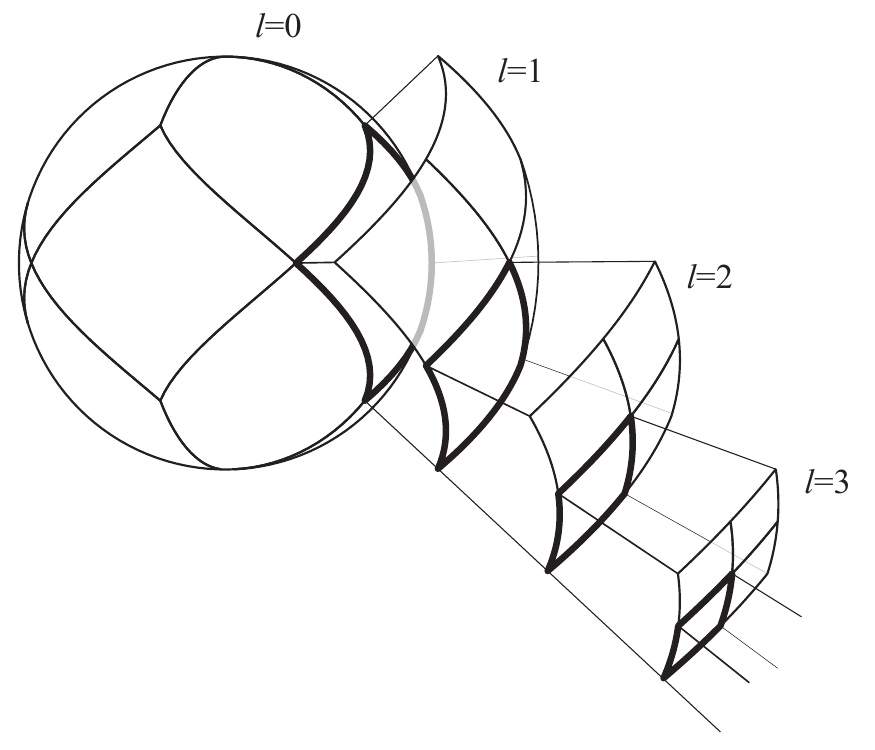}
    \caption{Illustration of the nested nature of \ac{HEALPix}. Level 0 divides the unit sphere into 12 equal-area tiles. In each subsequent level, every tile is divided equally into four new tiles.}
    \label{fig:healpix-levels}
\end{figure}

\ac{HEALPix} may be thought of as a tree in which each node except for the root node has four children (see Fig.~\ref{fig:healpix-levels}). At the lowest level in the tree, $l=0$, there are 12 base tiles, assigned integer indices $i = 0, 1, \dots, 11$. At level $l=1$, each of the 12 base tiles is subdivided into 4 new tiles. Every subsequent level divides each of the preceding level's tiles into 4 new tiles. At a given level $l$, each of the base tiles has been divided into $4^l$ tiles, i.e., $n_\mathrm{side} = 2^l$ pixels on each side. Thus there are $n_\mathrm{pix} = 12 (4^l) = 12(n_\mathrm{side})^2$ pixels at a given resolution, assigned integer indices from $i = 0, 1, \dots, n_\mathrm{pix}-1$.

The angular size of \ac{HEALPix} pixels varies from 59\arcdeg at $l=0$, all the way down to 0.39~mas at $l=29$, the highest level at which the pixel index can be stored as a 64 bit signed integer without overflow.

\subsection{RING and NESTED Ordering}

\begin{figure*}
    \includegraphics[width=\textwidth]{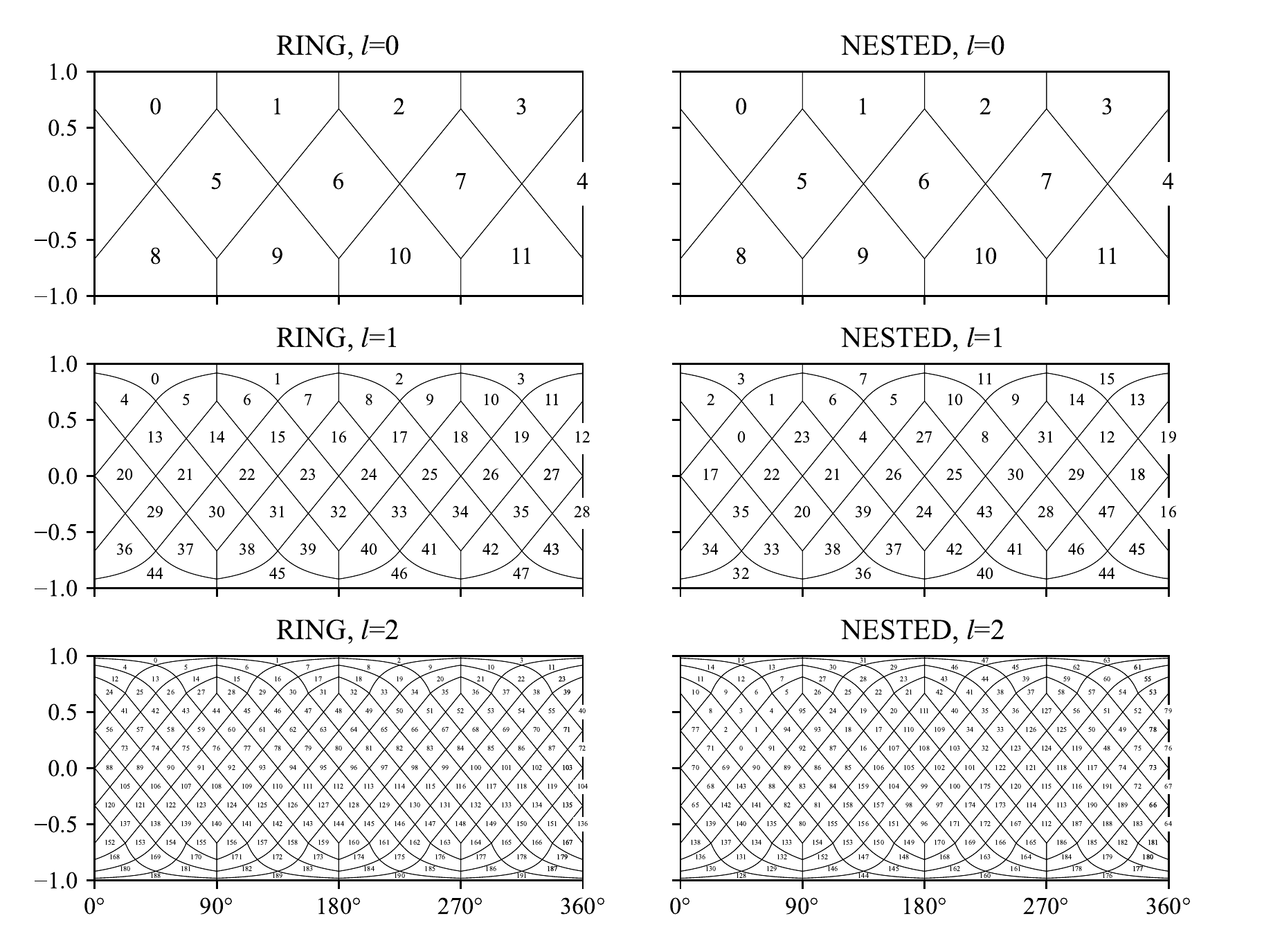}
    \caption{The first three levels of the \ac{HEALPix} RING and NESTED indexing schemes. In each panel, the horizontal axis is R.A., and the vertical axis is the sine of the decl. After \citet{2005ApJ...622..759G}.}
    \label{fig:ring-nested}
\end{figure*}

There are two conventional \ac{HEALPix} pixel-ordering schemes, called RING and NESTED (see Fig.~\ref{fig:ring-nested}). At level $l=0$, the two ordering schemes are identical, but they differ at all higher orders. In the RING scheme, the pixel index $i$ advances first with R.A. from west to east and then with decl. from north to south. In the NESTED scheme, pixels that are siblings of one another in the \ac{HEALPix} tree have consecutive values of pixel index.

Thus a \ac{HEALPix} tile at any resolution is fully specified by a tuple of three values: the indexing scheme (RING or NESTED), the resolution level $l$ (or equivalently, $n_\mathrm{side}$), and the pixel index $i$. \ac{HEALPix} software libraries typically provide two functions, one to convert from pixel index to R.A. and decl., and one to do the reverse. In Healpy \citep{2019JOSS....4.1298Z} these are called \lstinline{pix2ang} and \lstinline{ang2pix} respectively. In astropy-healpix \citep{2020ascl.soft11023R}, these are called \lstinline{healpix_to_lonlat} and \lstinline{lonlat_to_healpix}.

The NESTED scheme has the delightful property that the base 4 digits of the pixel index $i$ encode the path all the way from the root of the \ac{HEALPix} tree to the leaf tile. We may write a pixel index at any level $l^*$ in the mixed-radix form
$$
i = \underbrace{(i_0)_{12}}_{l=0} \underbrace{(i_1)_4}_{l=1} \underbrace{(i_2)_4}_{l=2} \dots \underbrace{(i_{l^*})_4}_{l=l^*},
$$
which expands to
$$
i = 4^{(l^*)} i_0
  + 4^{(l^* - 1)} i_1
  + 4^{(l^* - 2)} i_2
  + \dots
  + i_{l^*}.
$$

There is yet a third pixel encoding called UNIQ \citep{2015A&A...580A.132R}, which packs the resolution and the NESTED pixel index into a single integer:
$$
u = i + 4^{l + 1} = i + 4 (n_\mathrm{side})^2.
$$
The pixel index and resolution can be recovered from the UNIQ representation using bitwise operations.

\subsection{HEALPix Image and Region Formats}

Conventionally, the in-memory or on-disk (e.g., FITS file format; \citealt{ 2010A&A...524A..42P}) representation of an all-sky \ac{HEALPix} image is a 1D array of length $n_\mathrm{pix}$. The $i$th value of the array is simply the value of the image sampled at the center of the pixel (or perhaps the value of the image integrated over the area of the pixel, depending on the application) with pixel index $i$. The ordering (RING or NESTED) and the uniform \ac{HEALPix} resolution $n_\mathrm{nside}$ are stored as metadata. This flat-resolution format is prevalent in \ac{CMB} applications and dust maps, and up through \ac{O2} was used as the native format for \ac{GW} probability maps.

\begin{figure*}
    \includegraphics[width=\textwidth]{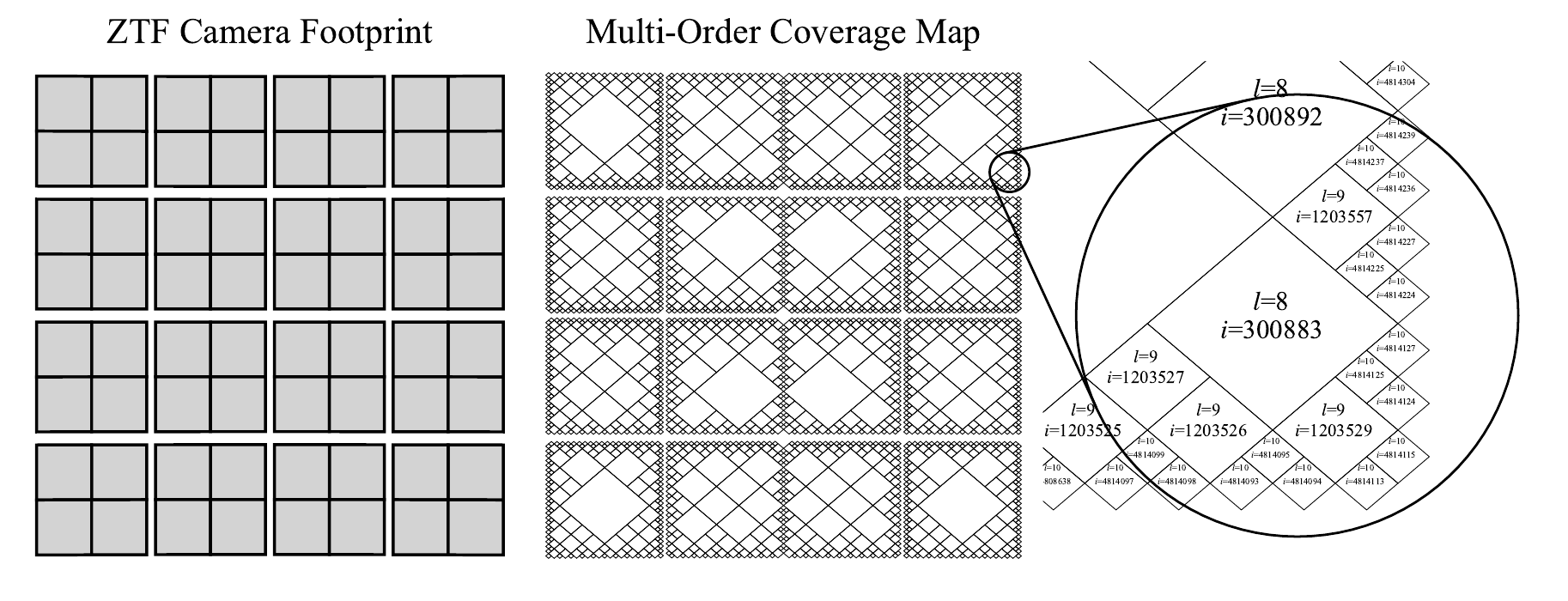}
    \caption{The 47~deg$^2$ footprint of the \ac{ZTF} camera as a \ac{HEALPix} \ac{MOC}, refined to a maximum level of $l=10$. The left panel shows the footprint as a collection of filled polygons: each of the small filled squares is one \ac{CCD} quadrant, each $2 \times 2$ cluster of quadrants comprise a \ac{CCD}, and the $4 \times 4$ grid of \acp{CCD} comprise the entire focal plane. The middle panel shows the outlines of the multi-resolution \ac{HEALPix} tiles that comprise the \ac{MOC}. The inset panel at right shows a small portion of the \ac{MOC}, with each pixel labeled with its resolution level $l$ and its nested pixel index $i$.}
    \label{fig:ztf-footprint}
\end{figure*}

It is also possible to store a region on the sphere---for example, the footprint of an observation or of a survey---as a set of \ac{HEALPix} pixels. Fig.~\ref{fig:ztf-footprint} shows the footprint of the 47~deg$^2$ \ac{ZTF} camera as a \ac{MOC}, consisting of a list of \ac{HEALPix} tiles of mixed resolutions that are inside the region. The on-disk representation of a \ac{MOC} in the \ac{FITS} format is simply a 1D array of UNIQ indices. A \ac{MOC} is typically a much more compact representation than a flat, fixed-resolution pixel mask: much of the interior of the region can be encoded using a small number of low-resolution tiles, and high-resolution tiles are only needed near the region's boundary.

From \ac{O3} onward, the native format of \ac{GW} probability sky maps is a multi-resolution FITS format based on the \ac{MOC} format: it is table with one column containing the UNIQ pixel indices of each HEALPix tile, and additional columns containing floating-point values associated with each tile (see Fig.~\ref{fig:skymap-mesh}). Because \ac{GW} sky maps are generated using an adaptive \ac{HEALPix} mesh refinement scheme, devoting higher resolution to regions of higher probability density \citep{2016PhRvD..93b4013S}, the savings in memory is substantial.

\subsection{Range Sets}

A final representation of a set of multi-resolution \ac{HEALPix} tiles is as a range set. This is often the most computationally convenient form, and is the most important for this paper. One first selects a fixed maximum resolution level, typically $l_\mathrm{max} = 29$ because it is the highest level at which pixel indices can fit in signed 64 bit integers. (Although \emph{unsigned} 64 bit integers can hold $l_\mathrm{max} = 30$ pixel indices, they are seldom used because most \ac{HEALPix} libraries use the pixel index -1 to represent error conditions.) Now observe that a given \ac{HEALPix} tile of level $l$ and pixel index $i$ contains all of the descendant pixels at level $l_\mathrm{max}$, such that their indices $i_\mathrm{max}$ are in the right-half-open interval,
$$
    \left[4^{(l_\mathrm{max} - l)} i,\, 4^{(l_\mathrm{max} - l)} (i + 1)\right).
$$
Each tile in a \ac{MOC} may be described by such an interval, and the \ac{MOC} as a whole can be described by the integer set consisting of a union of disjoint intervals. That collection is called an interval set or a range set. Table~\ref{tab:rangeset} lists the pixels that are visible within the inset panel of Fig~\ref{fig:ztf-footprint} as a range set.

\ac{HEALPix} range sets are advantageous because the complicated problem of combining (e.g., taking the union or intersection of) multiple regions and merging overlapping tiles simplifies to the easier problem of combining sets of integer ranges. There is a straightforward algorithm to merge any number of range sets. In pseudocode:
\begin{enumerate}
    \item \label{item:cat}Concatenate all of the range sets into a single list of ranges.
    \item \label{item:sort}Sort the list of ranges by their lower bounds, breaking ties by their upper bounds.
    \item \label{item:traverse}Walk through the list element-by-element and merge overlapping endpoints.
\end{enumerate}

Some \ac{MOC} implementations like MOCPy provide only a binary union operator to merge a pair of range sets; if there are $k$ range sets and a total of $n$ ranges, then applying the above algorithm pairwise and recursively costs $O(n k \log n)$ time. However, the algorithm as written above supports an arbitrary number of ranges sets; it is dominated by the sort in Step~\ref{item:sort} and costs only $O(n \log n)$ time. If there are $k$ range sets and they are presorted, then Steps~\ref{item:cat}~and~\ref{item:sort} can be replaced by a $k$-way merge, and the algorithm completes in $O(n \log k)$ time \citep{mehlhorn2008algorithms}. If all of the intervals in all of the range sets are stored in a single presorted list to begin with (if, for example, they are stored in a single table in a relational database, with different range sets distinguished by a foreign key), then the element-by-element traversal in Step~\ref{item:traverse} dominates, and the algorithm completes in only $O(n)$ time.

This is a crucial point: we can evaluate unions of large numbers of regions ($k \gtrsim 10$) orders of magnitude faster if we store all of the ranges of all of the range sets in a single sorted data structure, in one table of a PostgreSQL database. The speedup is remarkable in the following example. There are $k=1830$ standard \ac{ZTF} fields. The \ac{ZTF} footprint with quadrant-level detail in Fig.~\ref{fig:ztf-footprint} contains 826 HEALPix tiles down to $l=10$, for a total of $n \approx 1830 \times 826 = 1{,}511{,}580$ HEALPix tiles. In this example, the database approach requires $k \log_2 n \approx 4 \times 10^4$ times fewer comparisons than the naive pairwise union approach, $\log_2 n \approx 20$ times fewer comparisons than a naive $k$-way union approach, and $\log_2 n / \log_2 k \approx 2$ times fewer comparisons than an algorithm that does a $k$-way sorted merge.

There is also an efficient algorithm to test if a point is within a \ac{MOC}, provided the range set is presorted. Calculate the level $l_\mathrm{max}$ pixel index of the point, then simply perform a bisection search to find a matching interval (or no matching interval) in $O(\log n)$ time.

\begin{deluxetable}{cccc}
\tablecaption{Range Set Example for the Inset from Fig.~\ref{fig:ztf-footprint}\label{tab:rangeset}}
\tablehead{
    \colhead{$l$} &
    \colhead{$i$} &
    \colhead{UNIQ} &
    \colhead{Range Set (at $l = l_\mathrm{max} = 29$)}
}
\startdata
\vdots & \vdots &\vdots & \vdots \\
8 & 300883 & 563027 & [1323297428400504832, 1323301826447015936) \\
8 & 300892 & 563036 & [1323337010819104768, 1323341408865615872) \\
\vdots & \vdots &\vdots & \vdots \\
9 & 1203525 & 2252101 & [1323289731819110400, 1323290831330738176) \\
9 & 1203526 & 2252102 & [1323290831330738176, 1323291930842365952) \\
9 & 1203527 & 2252103 & [1323291930842365952, 1323293030353993728) \\
9 & 1203529 & 2252105 & [1323294129865621504, 1323295229377249280) \\
9 & 1203557 & 2252133 & [1323324916191199232, 1323326015702827008) \\
\vdots & \vdots &\vdots & \vdots \\
10 & 4808638 & 9002942 & [1321788348691382272, 1321788623569289216) \\
\vdots & \vdots &\vdots & \vdots \\
10 & 4814093 & 9008397 & [1323287807673761792, 1323288082551668736) \\
10 & 4814094 & 9008398 & [1323288082551668736, 1323288357429575680) \\
10 & 4814095 & 9008399 & [1323288357429575680, 1323288632307482624) \\
10 & 4814097 & 9008401 & [1323288907185389568, 1323289182063296512) \\
10 & 4814098 & 9008402 & [1323289182063296512, 1323289456941203456) \\
10 & 4814099 & 9008403 & [1323289456941203456, 1323289731819110400) \\
10 & 4814113 & 9008417 & [1323293305231900672, 1323293580109807616) \\
10 & 4814115 & 9008419 & [1323293854987714560, 1323294129865621504) \\
10 & 4814124 & 9008428 & [1323296328888877056, 1323296603766784000) \\
10 & 4814125 & 9008429 & [1323296603766784000, 1323296878644690944) \\
10 & 4814127 & 9008431 & [1323297153522597888, 1323297428400504832) \\
10 & 4814224 & 9008528 & [1323323816679571456, 1323324091557478400) \\
10 & 4814225 & 9008529 & [1323324091557478400, 1323324366435385344) \\
10 & 4814227 & 9008531 & [1323324641313292288, 1323324916191199232) \\
10 & 4814236 & 9008540 & [1323327115214454784, 1323327390092361728) \\
10 & 4814237 & 9008541 & [1323327390092361728, 1323327664970268672) \\
10 & 4814239 & 9008543 & [1323327939848175616, 1323328214726082560) \\
10 & 4814304 & 9008608 & [1323345806912126976, 1323346081790033920) \\
\vdots & \vdots &\vdots & \vdots \\
\enddata
\end{deluxetable}

\section{PostgreSQL and Multiranges}
\label{sec:postgresql}

PostgreSQL \citep{PostgreSQL} is an established, popular, open-source, relational database management system with a \ac{SQL} interface. It is repackaged and sold by a variety of cloud providers as part of their flagship managed database services in Amazon \ac{RDS}, Google Cloud SQL, Azure Database, and so on. It is a common choice of database for back ends of web applications, especially science applications and particularly astronomy brokers, \acp{TOM}, and marshals.

One of many reasons for PostgreSQL's popularity in the sciences is its wide variety of built-in data types. It may come as no surprise that PostgreSQL has supported range types since version 9.2.0, released in 2012. Specifically, the \lstinline{INT8RANGE} type is ideal for storing ranges of 64 bit, 8 byte, $l_\mathrm{max}=29$ \ac{HEALPix} indices, as described in the previous section. PostgreSQL defines many Boolean comparison operations on ranges: it can test if one range contains another, if one range overlaps another, if a range contains a scalar, etc. These operations are accelerated on columns that are indexed using the \acl{GiST} (\acsu{GiST}; \citealt{GiST}) or \acl{SP-GiST} (\acsu{SP-GiST}; \citealt{SP-GiST}) methods.

PostgreSQL 14.0, released on 2021 September 30, added two features that together make it possible to perform \ac{HEALPix} \ac{MOC} queries directly within the database. The first feature is a new \emph{multirange} type, consisting of an array of ranges. The \lstinline{INT8MULTIRANGE} type corresponds to \ac{HEALPix} range sets described in the previous section. The second feature is the \lstinline{range_agg} aggregate function, which takes ranges as its input and returns their union as a multirange.

\section{HEALPix Alchemy}
\label{sec:api}

In all but the simplest web applications, it is common to generate database queries using a high-level abstraction layer rather than issuing hard-coded query statements directly. One of the more popular database abstraction libraries for Python is SQLAlchemy \citep{sqlalchemy}, which allows one to express queries using Python syntax and provides a degree of independence between the code and the choice of database engine.

We have written HEALPix Alchemy, a Python package that extends SQLAlchemy to make it easy to work with multi-resolution \ac{HEALPix} geometry. The match between \ac{HEALPix} range sets and PostgreSQL multiranges is so perfect that HEALPix Alchemy consists of barely 100 lines of code (as measured using cloc; \citealt{adanial_cloc}). We summarize the design of HEALPix Alchemy below.

\subsection{Column Types}

HEALPix Alchemy adds two custom SQLAlchemy column types (\emph{type decorators}) that wrap the database's own built-in \ac{SQL} types: \lstinline{healpix_alchemy.Point} and \lstinline{healpix_alchemy.Tile}. For both column types, if a column is declared with the \lstinline{index=True} keyword argument, HEALPix Alchemy automatically selects the \ac{SP-GiST} indexing method for that column.

\subsubsection{The \lstinline{healpix_alchemy.Point} Class}

This class represents an infinitesimal point with no area, stored as a \ac{HEALPix} NESTED pixel index at $l=l_\mathrm{max}=29$. A table containing a column of this type could hold a catalog of distant galaxies or a list of optical transients. It maps to the built-in PostgreSQL \lstinline{BIGINT} type, a signed 64 bit integer. Values for \lstinline{healpix_alchemy.Point} columns can be initialized from any of the following Python objects:
\begin{itemize}
    \item an instance of \lstinline{astropy.coordinates.SkyCoord};
    \item a sequence of two \lstinline{astropy.units.Quantity} instances with angle units, which will be interpreted as the R.A. and decl. of the point; or
    \item an integer representing the \ac{HEALPix} NESTED index of the point at $l=l_\mathrm{max}=29$.
\end{itemize}

\subsubsection{The \lstinline{healpix_alchemy.Tile} Class}

This class represents a multi-resolution \ac{HEALPix} tile with finite area stored as a right-half-open interval of \ac{HEALPix} NESTED pixel indices at $l=l_\mathrm{max}=29$. It maps to the built-in PostgreSQL \lstinline{INT8RANGE} type, which is the range type corresponding to \lstinline{BIGINT}. A table containing a column of this type could store \acp{MOC} or \ac{GW} probability maps. Values for \lstinline{healpix_alchemy.Tile} columns can be initialized from any of the following Python objects:
\begin{itemize}
\item a single integer that will be interpreted as the address of the tile in the UNIQ indexing scheme;
\item a sequence of two integers like \lstinline{(1234, 5678)}, which will be interpreted as the lower and upper bounds of the right-half-open pixel index interval at $l=l_\mathrm{max}=29$; or
\item a string like \lstinline{'[1234,5678)'}, which is the canonical string representation of an \lstinline{INT8RANGE} in PostgreSQL.
\end{itemize}
There is also a factory method \lstinline{healpix_alchemy.Tile.tiles_from} that returns a collection of Python values suitable for initializing multiple \lstinline{healpix_alchemy.Tile} values. It accepts either of the following Python objects:
\begin{itemize}
    \item an instance of \lstinline{astropy.coordinates.SkyCoord} containing a vector of coordinates representing the vertices of a spherical polygon, which is converted to a \ac{MOC} with a default refinement level of $l=10$ using MOCPy \citep{2019ASPC..521..487B}; or
    \item an instance of \lstinline{mocpy.MOC}.
\end{itemize}

The \lstinline{healpix_alchemy.Tile} class provides the following properties:
\begin{itemize}
    \item \lstinline{healpix_alchemy.Tile.lower}, returning the left (closed) bound of the interval;
    \item \lstinline{healpix_alchemy.Tile.upper}, returning the right (open) bound of the interval;
    \item \lstinline{healpix_alchemy.Tile.length}, returning the difference of the right and left bounds; and
    \item \lstinline{healpix_alchemy.Tile.area}, returning the area of the tile in steradians.
\end{itemize}

Note that we do not provide a type decorator for the \lstinline{INT8MULTIRANGE} type itself because in most applications it should be more efficient to store ranges rather than multiranges in tables.

\subsection{Aggregate Functions}

HEALPix Alchemy provides the function \lstinline{healpix_alchemy.func.union} to find the union of \ac{HEALPix} ranges. Because it involves a \ac{SQL} aggregate function, generally it should be used in a subquery (examples to follow). The Python expression \lstinline{healpix_alchemy.func.union(x)} maps to the \ac{SQL} expression \lstinline{unnest(range_agg(x))}.

\section{Sample Code}
\label{sec:tutorial}

In this section, we provide some Python sample code using HEALPix Alchemy to perform the most common queries that one needs in a multimessenger astronomy broker or marshal.

\subsection{Installation}

HEALPix Alchemy requires Python 3.7 or later. To install HEALPix Alchemy and all of its Python dependencies from the Python Package Index using the \lstinline{pip} package manager, simply run the following command in a terminal:
\begin{lstlisting}
pip install healpix-alchemy
\end{lstlisting}

\subsection{Imports and Setup}

We begin with some imports:
\begin{lstlisting}
import sqlalchemy as sa
from sqlalchemy.ext.declarative import (
    as_declarative, declared_attr)
import healpix_alchemy as ha
\end{lstlisting}

SQLAlchemy needs to know the name for each table. You could provide the name by setting the \lstinline{__tablename__} attribute in each Python model class, but it is common practice to create a base class that generates the table name automatically from the Python class name:
\begin{lstlisting}
@as_declarative()
class Base:

    @declared_attr
    def __tablename__(cls):
        return cls.__name__.lower()
\end{lstlisting}

\subsection{Model Classes}

Next, we declare Python classes that will correspond to tables in the database. Each row of the \lstinline{Galaxy} table represents a point in a galaxy catalog:
\begin{lstlisting}
class Galaxy(Base):
    id = sa.Column(
        sa.Text, primary_key=True)
    hpx = sa.Column(
        ha.Point, index=True, nullable=False)
\end{lstlisting}
Each row of the \lstinline{Field} table represents the footprint of a \ac{ZTF} standard field:
\begin{lstlisting}
class Field(Base):
    id = sa.Column(
        sa.Integer, primary_key=True)
    tiles = sa.orm.relationship(
        lambda: FieldTile)
\end{lstlisting}
Each row of the \lstinline{FieldTile} table represents a multi-resolution \ac{HEALPix} tile that is contained within the corresponding field. There is a one-to-many mapping between \lstinline{Field} and \lstinline{FieldTile}.
\begin{lstlisting}
class FieldTile(Base):
    id = sa.Column(
        sa.ForeignKey(Field.id), primary_key=True)
    hpx = sa.Column(
        ha.Tile, primary_key=True, index=True)
\end{lstlisting}
Each row of the \lstinline{Skymap} table represents a \ac{GW} \ac{HEALPix} localization map:
\begin{lstlisting}
class Skymap(Base):
    id = sa.Column(
        sa.Integer, primary_key=True)
    tiles = sa.orm.relationship(
        lambda: SkymapTile)
\end{lstlisting}
Each row of the \lstinline{SkymapTile} table represents a multi-resolution \ac{HEALPix} tile with an associated probability density within a \ac{GW} localization map. There is a one-to-many mapping between \lstinline{Skymap} and \lstinline{SkymapTile}.
\begin{lstlisting}
class SkymapTile(Base):
    id = sa.Column(
        sa.ForeignKey(Skymap.id),
        primary_key=True)
    hpx = sa.Column(
        ha.Tile, primary_key=True, index=True)
    probdensity = sa.Column(
        sa.Float, nullable=False)
\end{lstlisting}

Finally, connect to the database, create all the tables, and start a session (replacing \lstinline{user}, \lstinline{password}, \lstinline{host}, and \lstinline{database} with the username, password, hostname, and database name respectively that you use to connect to your PostgreSQL database:
\begin{lstlisting}
url = 'postgresql://user:password@host/database'
engine = sa.create_engine(url)
Base.metadata.create_all(engine)
session = sa.orm.Session(engine)
\end{lstlisting}

\subsection{Populate with Sample Data}
Now we populate the tables with some sample data. First, we load the \ac{2MASS} Redshift Survey \citep{2012ApJS..199...26H} into the \lstinline{Galaxy} table. This catalog contains 44{,}599 galaxies. (It may take up to a minute for this to finish. Advanced users may speed this up significantly by vectorizing the conversion from \lstinline{SkyCoord} to \ac{HEALPix} indices and using SQLAlchemy bulk insertion.)
\begin{lstlisting}
from astropy.coordinates import SkyCoord
from astroquery.vizier import Vizier

catalog_name = 'J/ApJS/199/26/table3'
columns = ['SimbadName', 'RAJ2000', 'DEJ2000']
vizier = Vizier(columns=columns, row_limit=-1)
cat, = vizier.get_catalogs(catalog_name)

coords = SkyCoord(cat['RAJ2000'], cat['DEJ2000'])

for name, coord in zip(cat['SimbadName'], coords):
    session.add(Galaxy(id=name, hpx=coord))

session.commit()
\end{lstlisting}
Next, we load the footprints of the \ac{ZTF} standard fields into the \lstinline{Field} and \lstinline{FieldTile} tables. (It may take up to a minute for this to finish too. Advanced users may speed this up significantly by using SQLAlchemy bulk insertion.)
\begin{lstlisting}
from astropy.table import Table
from astropy.coordinates import SkyCoord
from astropy import units as u

url = 'https://raw.githubusercontent.com/ZwickyTransientFacility/ztf_information/master/field_grid/ztf_field_corners.csv'

for row in Table.read(url):
    id = int(row['field'])
    ras = row['ra1', 'ra2', 'ra3', 'ra4']
    decs = row['dec1', 'dec2', 'dec3', 'dec4']
    corners = SkyCoord(ras, decs, unit=u.deg)
    tiles = [
        FieldTile(hpx=hpx)
        for hpx in ha.Tile.tiles_from(corners)]
    session.add(Field(id=id, tiles=tiles))

session.commit()
\end{lstlisting}
Lastly, we load a sky map for LIGO/Virgo event GW200115\_042309 (\citealt{2021ApJ...915L...5A}; GraceDB ID S200115j) into the \lstinline{Skymap} and \lstinline{SkymapTile} tables.
\begin{lstlisting}
url = 'https://gracedb.ligo.org/apiweb/superevents/S200115j/files/bayestar.multiorder.fits'
data = Table.read(url)

tiles = [
    SkymapTile(
        hpx=row['UNIQ'],
        probdensity=row['PROBDENSITY'])
    for row in data]

session.add(Skymap(id=1, tiles=tiles))
session.commit()
\end{lstlisting}

\subsection{Example Queries}

Now we provide some examples of common queries that would occur in a multimessenger astronomy broker or marshal. (In all of the examples below, we limit the result set to 5 rows to avoid generating a large amount of terminal output.)

\subsubsection{What Is the Area of Each Field?}

The area of a region is simply the sum of the area of all of the \ac{HEALPix} tiles that belong to the region. The following query:
\begin{lstlisting}
query = session.query(
    FieldTile.id, sa.func.sum(FieldTile.hpx.area)
).group_by(
    FieldTile.id
).limit(
    5
)

for id, area in session.execute(query):
    print(f'Field {id} has area {area:.3g} sr')
\end{lstlisting}
produces this output:
\begin{lstlisting}
Field 199 has area 0.0174 sr
Field 200 has area 0.0174 sr
Field 201 has area 0.0174 sr
Field 202 has area 0.0174 sr
Field 203 has area 0.0174 sr
\end{lstlisting}

\subsubsection{How Many Galaxies Are in Each Field?}

For this query, we need to introduce the \lstinline{contains} comparison function, which tests if a \lstinline{healpix_alchemy.Tile} contains a \lstinline{healpix_alchemy.Point}. Behind the scenes, this simply maps to the built-in PostgreSQL \lstinline{@>} comparison operator. The following query:
\begin{lstlisting}
count = sa.func.count(Galaxy.id)

query = session.query(
    FieldTile.id, count
).filter(
    FieldTile.hpx.contains(Galaxy.hpx)
).group_by(
    FieldTile.id
).order_by(
    count.desc()
).limit(
    5
)

for id, n in session.execute(query):
    print(f'Field {id} contains {n} galaxies')
\end{lstlisting}
produces this output:
\begin{lstlisting}
Field 1739 contains 343 galaxies
Field 699 contains 336 galaxies
Field 700 contains 311 galaxies
Field 225 contains 303 galaxies
Field 1740 contains 289 galaxies
\end{lstlisting}

\subsubsection{What Is the Probability Density at the Position of Each Galaxy?}

Since sky map tiles and region tiles are represented using the same \lstinline{healpix_alchemy.Tile} column type; this is just a minor variation on the previous query. The following:
\begin{lstlisting}
query = session.query(
    Galaxy.id, SkymapTile.probdensity
).filter(
    SkymapTile.id == 1,
    SkymapTile.hpx.contains(Galaxy.hpx)
).order_by(
    SkymapTile.probdensity.desc()
).limit(
    5
)

for id, p in session.execute(query):
    print(f'{id} has prob. density {p:.5g}/sr')
\end{lstlisting}
produces this output:
\begin{lstlisting}
2MASX J02532153+0632222 has prob. density 20.701/sr
2MASX J02530482+0555431 has prob. density 20.695/sr
2MASX J02533119+0628252 has prob. density 20.669/sr
2MASX J02524584+0639206 has prob. density 20.656/sr
2MASX J02534120+0615562 has prob. density 20.567/sr
\end{lstlisting}

\subsubsection{What Is the Probability Contained within Each Field?}

For this query, we need to introduce the \lstinline{overlaps} comparison function, which tests if one \lstinline{healpix_alchemy.Tile} overlaps another. Behind the scenes, this simply maps to the built-in PostgreSQL \lstinline{&&} comparison operator. We also need the \lstinline{*} operator, which returns a new tile that is the intersection of two tiles. The following query:
\begin{lstlisting}
area = (FieldTile.hpx * SkymapTile.hpx).area
prob = sa.func.sum(SkymapTile.probdensity * area)

query = session.query(
    FieldTile.id, prob
).filter(
    SkymapTile.id == 1,
    FieldTile.hpx.overlaps(SkymapTile.hpx)
).group_by(
    FieldTile.id
).order_by(
    prob.desc()
).limit(
    5
)

for id, prob in session.execute(query):
    print(f'Field {id} probability is {prob:.3g}')
\end{lstlisting}
produces this output:
\begin{lstlisting}
Field 1499 probability is 0.165
Field 1446 probability is 0.156
Field 452 probability is 0.154
Field 505 probability is 0.0991
Field 401 probability is 0.0962
\end{lstlisting}

\subsubsection{What Is the Combined Area of Fields 1000 through 2000?}

In the next two examples, we introduce \lstinline{healpix_alchemy.func.union()}, which
finds the union of a set of tiles. Because it is an aggregate function, it
should generally be used in a subquery. The following:
\begin{lstlisting}
union = session.query(
    ha.func.union(FieldTile.hpx).label('hpx')
).filter(
    FieldTile.id.between(1000, 2000)
).subquery()

query = session.query(
    sa.func.sum(union.columns.hpx.area)
)

result = session.execute(query).scalar_one()
print(f'{result:.3g} sr')
\end{lstlisting}
produces this output:
\begin{lstlisting}
9.33 sr
\end{lstlisting}

\subsubsection{What Is the Integrated Probability Contained within Fields 1000 through 2000?}

This is a minor variation on the previous query. The following:
\begin{lstlisting}
union = session.query(
    ha.func.union(FieldTile.hpx).label('hpx')
).filter(
    FieldTile.id.between(1000, 2000)
).subquery()

area = (union.columns.hpx * SkymapTile.hpx).area
prob = sa.func.sum(SkymapTile.probdensity * area)

query = session.query(
    prob
).filter(
    SkymapTile.id == 1,
    union.columns.hpx.overlaps(SkymapTile.hpx)
)

result = session.execute(query).scalar_one()
print(f'{result:.3g}')
\end{lstlisting}
produces this output:
\begin{lstlisting}
0.837
\end{lstlisting}

\subsubsection{What Is the Area of the 90\% Credible Region?}

The 90\% credible region of a \ac{GW} probability sky map is defined as the region with the smallest area that has a 90\% probability of containing the true location of the \ac{GW} source. To find the \ac{HEALPix} tiles that are within the 90\% credible region, we have to rank the tiles by descending probability density, then calculate the cumulative sum of their probability (probability density times area), and then search for the tile that has a cumulative probability of 0.9.

In \ac{SQL}, a cumulative is expressed as a window expression, using an \lstinline{over} clause. This query involves such a clause as well as a subquery. The following:
\begin{lstlisting}
# Assemble the expression representing the
# cumulative area of credible levels
cum_area = sa.func.sum(
    SkymapTile.hpx.area
).over(
    order_by=SkymapTile.probdensity.desc()
).label(
    'cum_area'
)

# Assemble the expression representing the
# cumulative probability of credible levels
cum_prob = sa.func.sum(
    SkymapTile.probdensity * SkymapTile.hpx.area
).over(
    order_by=SkymapTile.probdensity.desc()
).label(
    'cum_prob'
)

# Query cumulative area and probability of
# credible levels
subquery = session.query(
    cum_area, cum_prob
).filter(
    SkymapTile.id == 1
).subquery()

# Find cumulative area at which
# cumulative probability reaches 90%
query = session.query(
    sa.func.max(subquery.columns.cum_area)
).filter(
    subquery.columns.cum_prob <= 0.9
)

result = session.execute(query).scalar_one()
print(f'{result:.3g} sr')
\end{lstlisting}
produces this output:
\begin{lstlisting}
0.277 sr
\end{lstlisting}

\subsubsection{Which Galaxies Are within the 90\% Credible Region?}

This is just a minor variation on the previous query. The following:
\begin{lstlisting}
# Assemble the expression representing the
# cumulative probability of credible levels
cum_prob = sa.func.sum(
    SkymapTile.probdensity * SkymapTile.hpx.area
).over(
    order_by=SkymapTile.probdensity.desc()
).label(
    'cum_prob'
)

# Query probability density
# and cumulative probability
subquery = session.query(
    SkymapTile.probdensity, cum_prob
).filter(
    SkymapTile.id == 1
).subquery()

# Find probability density at which
# cumulative probablity reaches 90%
min_probdensity = session.query(
    sa.func.min(subquery.columns.probdensity)
).filter(
    subquery.columns.cum_prob <= 0.9
).scalar_subquery()

# Query galaxies that have probability density
# great than or equal to that on boundary of
# the 90% credible level
query = session.query(
    Galaxy.id
).filter(
    SkymapTile.id == 1,
    SkymapTile.hpx.contains(Galaxy.hpx),
    SkymapTile.probdensity >= min_probdensity
).limit(
    5
)

for galaxy_id, in session.execute(query):
    print(galaxy_id)
\end{lstlisting}
produces this output:
\begin{lstlisting}
2MASX J02424077-0000478
2MASX J02352772-0921216
2MASX J02273746-0109226
2MASX J02414523+0026354
2MASX J20095408-4822462
\end{lstlisting}

\section{Performance}
\label{sec:performance}

To measure the performance of HEALPix Alchemy, we populated a test database with a $N_\mathrm{gal}=40\mathrm{k}$ randomly and isotropically distributed galaxies, a randomly subdivided sky map with $N_\mathrm{skymap}=20\mathrm{k}$ HEALPix tiles, and $N_\mathrm{field}$=1--10k random and isotropically distributed square fields with the dimensions of the \ac{ZTF} camera.

We measured the performance of a few representative queries based on those in the previous section:
\begin{enumerate}
    \item\label{item:area-of-union}\emph{Find area of union}: Find the area in steradians of all $N_\mathrm{field}$ fields.
    \item\label{item:crossmatch-galaxies}\emph{Cross-match with 40k galaxies}: Count the number of galaxies within each of the $N_\mathrm{field}$ fields.
    \item\label{item:crossmatch-skymap}\emph{Find fields in 90\% cred. region}: Count how many of the $N_\mathrm{field}$ fields are within the 90\% credible region of the sky map.
\end{enumerate}
(Queries \ref{item:crossmatch-galaxies} and \ref{item:crossmatch-skymap} count the number of matching galaxies or fields rather than listing in order to limit the amount of output sent back to the client while still causing PostgreSQL to enumerate all of the matching rows.)

To gather benchmarks on \ac{AWS}, we configured a PostgreSQL 14 database in the \ac{RDS} Database Preview Environment on one db.m5.xlarge instance. We installed Ubuntu 20.04 LTS (``Focal Fossa'') and HEALPix Alchemy on an \ac{EC2} instance in the same availability zone as the database. Both the \ac{RDS} and the \ac{EC2} instances ran on Intel Xeon Platinum 8000 series CPUs and had 4 virtual cores and 16~GiB of memory \citep{aws-instance-types}.

For small numbers of fields, we find that HEALPix Alchemy queries are comparable in run time to equivalent Python code using MOCPy. However, as the number of fields grows into the hundreds or thousands, HEALPix Alchemy is orders of magnitude faster than MOCPy, because in the database the tiles belonging to all of the \acp{MOC} are indexed and presorted as a single table.

\begin{figure}
    \includegraphics[width=\columnwidth]{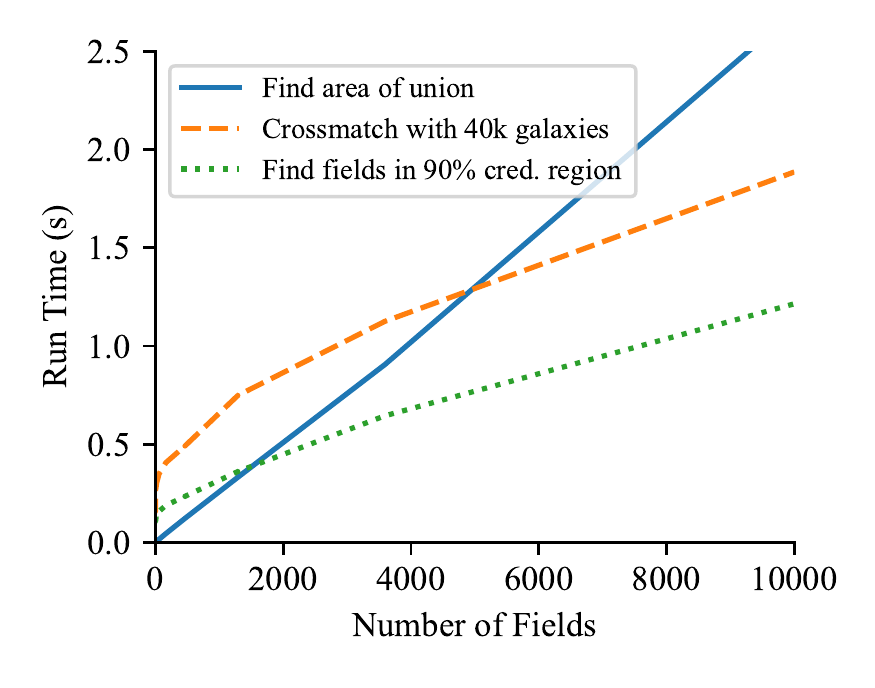}
    \caption{Run time on \ac{AWS} as a function of the number of fields of three representative queries in HEALPix Alchemy.}
    \label{fig:benchmark}
\end{figure}

Fig.~\ref{fig:benchmark} plots the run time in seconds of all three queries as a function of the number of fields, $N_\mathrm{field}$. Within the domain of this plot, all three queries exhibit roughly linear scaling. There are $N_\mathrm{field}=1830$ standard \ac{ZTF} fields. Reading off of the plot, for this database size all three queries complete within about a second---suitable for use in an interactive web application.

\section{Conclusion}

The new multirange type in PostgreSQL 14 is perfect for storing HEALPix range sets, facilitating spatial region and image queries that are the bread and butter of a multimessenger astronomy broker or marshal. We have provided the minimalist HEALPix Alchemy package to make it easier to express these queries in Python.

HEALPix Alchemy is built on the solid foundations of several existing high-quality open-source \ac{HEALPix} and \ac{MOC} implementations in Python. Specifically, HEALPix Alchemy integrates with Astropy and MOCPy to populate PostgreSQL tables from a variety of Python types. However, once spatial data are within the database, HEALPix Alchemy queries significantly outperform MOCPy for bulk operations on large numbers of \acp{MOC} because PostgreSQL can store all of the HEALPix tiles for all of the regions or sky maps in a single, coherently indexed table.

Our technique works with a stock PostgreSQL 14 server without any patches or extensions---an important feature because web applications often demand the robustness of a fully managed cloud database service. At the time of this writing, PostgreSQL 14 is supported by all three major cloud providers: Amazon \ac{RDS} \citep{aws-rds-postgresql-versions}, Google Cloud SQL \citep{google-cloud-postgresql-versions}, and (in some regions) Microsoft Azure Database \citep{postgresql-azure-14-shipped}.

HEALPix Alchemy provides spatial indexing for SkyPortal \citep{2019JOSS....4.1247V}, a general-purpose astronomical data portal, of which the \ac{ZTF} follow-up marshal, Fritz, is an instance. We are currently working on consolidating the multimessenger functionality of the GROWTH \ac{ToO} Marshal \citep{2021RMxAC..53...91A} into SkyPortal using HEALPix Alchemy for greatly improved performance. We expect that the HEALPix Alchemy technique will be widely applicable to science portals in the multimessenger astronomy era, including NASA's recently proposed Multimessenger Astrophysics Support Center \citep{2021arXiv210910841S}.

\begin{acknowledgments}
The open-source HEALPix Alchemy package is publicly available on GitHub (\url{https://github.com/skyportal/healpix-alchemy}) and Zenodo \citep{healpix-alchemy-1.0.1}.

L.P.S. acknowledges support for this work from an Internal Scientist Funding Model (ISFM) award from NASA for the Time-domain Astrophysics Coordination Hub (TACH) work package, a Science Task Group award from the Science and Exploration Directorate at NASA Goddard Space Flight Center, and partial support from the Neil Gehrels Swift Observatory project.
B.P. acknowledges support from a Northeastern Lawrence Co-op Fellowship.
M.W.C. acknowledges support from the National Science Foundation with grant Nos. PHY-2010970 and OAC-2117997.
J.S.B. was partially supported by a Gordon and Betty Moore Foundation Data-Driven Discovery grant. 
The authors thank Israel Martinez-Castellanos and Paul Jungwirth for feedback and suggestions on this paper.

Based on observations obtained with the Samuel Oschin 48 inch Telescope and the 60 inch Telescope at the Palomar Observatory as part of the Zwicky Transient Facility project. ZTF is supported by the National Science Foundation under grant No. AST-2034437 and a collaboration including Caltech, IPAC, the Weizmann Institute for Science, the Oskar Klein Center at Stockholm University, the University of Maryland, Deutsches Elektronen-Synchrotron and Humboldt University, the TANGO Consortium of Taiwan, the University of Wisconsin at Milwaukee, Trinity College Dublin, Lawrence Livermore National Laboratories, and IN2P3, France. Operations are conducted by COO, IPAC, and UW.

This document is LIGO-P2100433-v6.
\end{acknowledgments}

\software{%
Astropy \citep{2013A&A...558A..33A,2018AJ....156..123A},
astropy-healpix \citep{2020ascl.soft11023R},
Astroquery \citep{2019AJ....157...98G},
ligo.skymap \citep{2016PhRvD..93b4013S,2016ApJ...829L..15S},
Matplotlib \citep{2007CSE.....9...90H},
MOCPy \citep{2019ASPC..521..487B},
NumPy \citep{2020Natur.585..357H},
PostgreSQL \citep{PostgreSQL},
SkyPortal \citep{2019JOSS....4.1247V},
SQLAlchemy \citep{sqlalchemy},
\ac{HEALPix} \citep{2005ApJ...622..759G}.
}

\bibliography{autogenerated-references-do-not-edit-manually,references}{}

\begin{thebibliography}{}
\expandafter\ifx\csname natexlab\endcsname\relax\def\natexlab#1{#1}\fi
\providecommand{\url}[1]{\href{#1}{#1}}
\providecommand{\dodoi}[1]{doi:~\href{http://doi.org/#1}{\nolinkurl{#1}}}
\providecommand{\doeprint}[1]{\href{http://ascl.net/#1}{\nolinkurl{http://ascl.net/#1}}}
\providecommand{\doarXiv}[1]{\href{https://arxiv.org/abs/#1}{\nolinkurl{https://arxiv.org/abs/#1}}}

\bibitem[{{Aartsen} {et~al.}(2017){Aartsen}, {Ackermann}, {Adams}, {Aguilar},
  {Ahlers}, {Ahrens}, {Altmann}, {Andeen}, {Anderson}, {Ansseau}, \&
  et~al.}]{2017APh....92...30A}
{Aartsen}, M.~G., {Ackermann}, M., {Adams}, J., {et~al.} 2017, Astroparticle
  Physics, 92, 30, \dodoi{10.1016/j.astropartphys.2017.05.002}

\bibitem[{{Abbott} {et~al.}(2018){Abbott}, {Abbott}, {Abbott}, {Abernathy},
  {Acernese}, {Ackley}, {Adams}, {Adams}, {Addesso}, {Adhikari}, \&
  et~al.}]{2018LRR....21....3A}
{Abbott}, B.~P., {Abbott}, R., {Abbott}, T.~D., {et~al.} 2018, Living Reviews
  in Relativity, 21, 3, \dodoi{10.1007/s41114-018-0012-9}

\bibitem[{{Abbott} {et~al.}(2021){Abbott}, {Abbott}, {Abraham}, {Acernese},
  {Ackley}, {Adams}, {Adams}, {Adhikari}, {Adya}, {Affeldt}, \&
  et~al.}]{2021ApJ...915L...5A}
{Abbott}, R., {Abbott}, T.~D., {Abraham}, S., {et~al.} 2021, \apjl, 915, L5,
  \dodoi{10.3847/2041-8213/ac082e}

\bibitem[{{Acernese} {et~al.}(2015){Acernese}, {Agathos}, {Agatsuma}, {Aisa},
  {Allemandou}, {Allocca}, {Amarni}, {Astone}, {Balestri}, {Ballardin}, \&
  et~al.}]{2015CQGra..32b4001A}
{Acernese}, F., {Agathos}, M., {Agatsuma}, K., {et~al.} 2015, Classical and
  Quantum Gravity, 32, 024001, \dodoi{10.1088/0264-9381/32/2/024001}

\bibitem[{{Akutsu} {et~al.}(2021){Akutsu}, {Ando}, {Arai}, {Arai}, {Araki},
  {Araya}, {Aritomi}, {Aso}, {Bae}, {Bae}, {Baiotti}, {Bajpai}, {Barton},
  {Cannon}, {Capocasa}, {Chan}, {Chen}, {Chen}, {Chen}, {Chu}, {Chu}, {Eguchi},
  {Enomoto}, {Flaminio}, {Fujii}, {Fukunaga}, {Fukushima}, {Ge}, {Hagiwara},
  {Haino}, {Hasegawa}, {Hayakawa}, {Hayama}, {Himemoto}, {Hiranuma}, {Hirata},
  {Hirose}, {Hong}, {Hsieh}, {Huang}, {Huang}, {Huang}, {Ikenoue}, {Imam},
  {Inayoshi}, {Inoue}, {Ioka}, {Itoh}, {Izumi}, {Jung}, {Jung}, {Kajita},
  {Kamiizumi}, {Kanda}, {Kang}, {Kawaguchi}, {Kawai}, {Kawasaki}, {Kim}, {Kim},
  {Kim}, {Kim}, {Kimura}, {Kita}, {Kitazawa}, {Kojima}, {Kokeyama}, {Komori},
  {Kong}, {Kotake}, {Kozakai}, {Kozu}, {Kumar}, {Kume}, {Kuo}, {Kuo},
  {Kuroyanagi}, {Kusayanagi}, {Kwak}, {Lee}, {Lee}, {Lee}, {Leonardi}, {Lin},
  {Lin}, {Lin}, {Liu}, {Luo}, {Marchio}, {Michimura}, {Mio}, {Miyakawa},
  {Miyamoto}, {Miyazaki}, {Miyo}, {Miyoki}, {Morisaki}, {Moriwaki}, {Nagano},
  {Nagano}, {Nakamura}, {Nakano}, {Nakano}, {Nakashima}, {Narikawa}, {Negishi},
  {Ni}, {Nishizawa}, {Obuchi}, {Ogaki}, {Oh}, {Oh}, {Ohashi}, {Ohishi},
  {Ohkawa}, {Okutomi}, {Oohara}, {Ooi}, {Oshino}, {Pan}, {Pang}, {Park},
  {Arellano}, {Pinto}, {Sago}, {Saito}, {Saito}, {Sakai}, {Sakai}, {Sakuno},
  {Sato}, {Sato}, {Sawada}, {Sekiguchi}, {Sekiguchi}, {Shibagaki}, {Shimizu},
  {Shimoda}, {Shimode}, {Shinkai}, {Shishido}, {Shoda}, {Somiya}, {Son},
  {Sotani}, {Sugimoto}, {Suzuki}, {Suzuki}, {Tagoshi}, {Takahashi},
  {Takahashi}, {Takamori}, {Takano}, {Takeda}, {Takeda}, {Tanaka}, {Tanaka},
  {Tanaka}, {Tanaka}, {Tanaka}, {Tanioka}, {Tapia San Martin}, {Telada},
  {Tomaru}, {Tomigami}, {Tomura}, {Travasso}, {Trozzo}, {Tsang}, {Tsubono},
  {Tsuchida}, {Tsuzuki}, {Tuyenbayev}, {Uchikata}, {Uchiyama}, {Ueda},
  {Uehara}, {Ueno}, {Ueshima}, {Uraguchi}, {Ushiba}, {van Putten}, {Vocca},
  {Wang}, {Wu}, {Wu}, {Wu}, {Xu}, {Yamada}, {Yamamoto}, {Yamamoto}, {Yamamoto},
  {Yokogawa}, {Yokoyama}, {Yokozawa}, {Yoshioka}, {Yuzurihara}, {Zeidler},
  {Zhao}, \& {Zhu}}]{2021PTEP.2021eA101A}
{Akutsu}, T., {Ando}, M., {Arai}, K., {et~al.} 2021, Progress of Theoretical
  and Experimental Physics, 2021, 05A101, \dodoi{10.1093/ptep/ptaa125}

\bibitem[{Alekseyenko \& Lee(2007)}]{10.1093/bioinformatics/btl647}
Alekseyenko, A.~V., \& Lee, C.~J. 2007, Bioinformatics, 23, 1386,
  \dodoi{10.1093/bioinformatics/btl647}

\bibitem[{Amazon(2021)}]{aws-instance-types}
Amazon. 2021, Amazon EC2 M5 Instances.
\newblock \url{https://aws.amazon.com/ec2/instance-types/m5/}

\bibitem[{Amazon(2022)}]{aws-rds-postgresql-versions}
---. 2022, PostgreSQL on Amazon RDS.
\newblock
  \url{https://web.archive.org/web/20220318061718/https://docs.aws.amazon.com/AmazonRDS/latest/UserGuide/CHAP_PostgreSQL.html}

\bibitem[{{Anand} {et~al.}(2021){Anand}, {Andreoni}, {Goldstein}, {Kasliwal},
  {Ahumada}, {Barnes}, {Bloom}, {Bulla}, {Cenko}, {Cooke}, {Coughlin},
  {Nugent}, \& {Singer}}]{2021RMxAC..53...91A}
{Anand}, S., {Andreoni}, I., {Goldstein}, D.~A., {et~al.} 2021, in Revista
  Mexicana de Astronomia y Astrofisica Conference Series, Vol.~53, Revista
  Mexicana de Astronomia y Astrofisica Conference Series, 91--99,
  \dodoi{10.22201/ia.14052059p.2021.53.20}

\bibitem[{Aref \& Ilyas(2001)}]{SP-GiST}
Aref, W., \& Ilyas, I. 2001, J. Intell. Inf. Syst., 17, 215,
  \dodoi{10.1023/A:1012809914301}

\bibitem[{{Astropy Collaboration} {et~al.}(2013){Astropy Collaboration},
  {Robitaille}, {Tollerud}, {Greenfield}, {Droettboom}, {Bray}, {Aldcroft},
  {Davis}, {Ginsburg}, {Price-Whelan}, {Kerzendorf}, {Conley}, {Crighton},
  {Barbary}, {Muna}, {Ferguson}, {Grollier}, {Parikh}, {Nair}, {Unther},
  {Deil}, {Woillez}, {Conseil}, {Kramer}, {Turner}, {Singer}, {Fox}, {Weaver},
  {Zabalza}, {Edwards}, {Azalee Bostroem}, {Burke}, {Casey}, {Crawford},
  {Dencheva}, {Ely}, {Jenness}, {Labrie}, {Lim}, {Pierfederici}, {Pontzen},
  {Ptak}, {Refsdal}, {Servillat}, \& {Streicher}}]{2013A&A...558A..33A}
{Astropy Collaboration}, {Robitaille}, T.~P., {Tollerud}, E.~J., {et~al.} 2013,
  \aap, 558, A33, \dodoi{10.1051/0004-6361/201322068}

\bibitem[{{Astropy Collaboration} {et~al.}(2018){Astropy Collaboration},
  {Price-Whelan}, {Sip{\H{o}}cz}, {G{\"u}nther}, {Lim}, {Crawford}, {Conseil},
  {Shupe}, {Craig}, {Dencheva}, {Ginsburg}, {VanderPlas}, {Bradley},
  {P{\'e}rez-Su{\'a}rez}, {de Val-Borro}, {Aldcroft}, {Cruz}, {Robitaille},
  {Tollerud}, {Ardelean}, {Babej}, {Bach}, {Bachetti}, {Bakanov}, {Bamford},
  {Barentsen}, {Barmby}, {Baumbach}, {Berry}, {Biscani}, {Boquien}, {Bostroem},
  {Bouma}, {Brammer}, {Bray}, {Breytenbach}, {Buddelmeijer}, {Burke},
  {Calderone}, {Cano Rodr{\'\i}guez}, {Cara}, {Cardoso}, {Cheedella}, {Copin},
  {Corrales}, {Crichton}, {D'Avella}, {Deil}, {Depagne}, {Dietrich}, {Donath},
  {Droettboom}, {Earl}, {Erben}, {Fabbro}, {Ferreira}, {Finethy}, {Fox},
  {Garrison}, {Gibbons}, {Goldstein}, {Gommers}, {Greco}, {Greenfield},
  {Groener}, {Grollier}, {Hagen}, {Hirst}, {Homeier}, {Horton}, {Hosseinzadeh},
  {Hu}, {Hunkeler}, {Ivezi{\'c}}, {Jain}, {Jenness}, {Kanarek}, {Kendrew},
  {Kern}, {Kerzendorf}, {Khvalko}, {King}, {Kirkby}, {Kulkarni}, {Kumar},
  {Lee}, {Lenz}, {Littlefair}, {Ma}, {Macleod}, {Mastropietro}, {McCully},
  {Montagnac}, {Morris}, {Mueller}, {Mumford}, {Muna}, {Murphy}, {Nelson},
  {Nguyen}, {Ninan}, {N{\"o}the}, {Ogaz}, {Oh}, {Parejko}, {Parley}, {Pascual},
  {Patil}, {Patil}, {Plunkett}, {Prochaska}, {Rastogi}, {Reddy Janga},
  {Sabater}, {Sakurikar}, {Seifert}, {Sherbert}, {Sherwood-Taylor}, {Shih},
  {Sick}, {Silbiger}, {Singanamalla}, {Singer}, {Sladen}, {Sooley},
  {Sornarajah}, {Streicher}, {Teuben}, {Thomas}, {Tremblay}, {Turner},
  {Terr{\'o}n}, {van Kerkwijk}, {de la Vega}, {Watkins}, {Weaver}, {Whitmore},
  {Woillez}, {Zabalza}, \& {Astropy Contributors}}]{2018AJ....156..123A}
{Astropy Collaboration}, {Price-Whelan}, A.~M., {Sip{\H{o}}cz}, B.~M., {et~al.}
  2018, \aj, 156, 123, \dodoi{10.3847/1538-3881/aabc4f}

\bibitem[{Bayer(2012)}]{sqlalchemy}
Bayer, M. 2012, in The Architecture of Open Source Applications Volume II:
  Structure, Scale, and a Few More Fearless Hacks, ed. A.~Brown \& G.~Wilson
  (aosabook.org).
\newblock \url{http://aosabook.org/en/sqlalchemy.html}

\bibitem[{{Bellm} {et~al.}(2019){Bellm}, {Kulkarni}, {Graham}, {Dekany},
  {Smith}, {Riddle}, {Masci}, {Helou}, {Prince}, {Adams}, {Barbarino},
  {Barlow}, {Bauer}, {Beck}, {Belicki}, {Biswas}, {Blagorodnova}, {Bodewits},
  {Bolin}, {Brinnel}, {Brooke}, {Bue}, {Bulla}, {Burruss}, {Cenko}, {Chang},
  {Connolly}, {Coughlin}, {Cromer}, {Cunningham}, {De}, {Delacroix}, {Desai},
  {Duev}, {Eadie}, {Farnham}, {Feeney}, {Feindt}, {Flynn}, {Franckowiak},
  {Frederick}, {Fremling}, {Gal-Yam}, {Gezari}, {Giomi}, {Goldstein},
  {Golkhou}, {Goobar}, {Groom}, {Hacopians}, {Hale}, {Henning}, {Ho}, {Hover},
  {Howell}, {Hung}, {Huppenkothen}, {Imel}, {Ip}, {Ivezi{\'c}}, {Jackson},
  {Jones}, {Juric}, {Kasliwal}, {Kaspi}, {Kaye}, {Kelley}, {Kowalski},
  {Kramer}, {Kupfer}, {Landry}, {Laher}, {Lee}, {Lin}, {Lin}, {Lunnan},
  {Giomi}, {Mahabal}, {Mao}, {Miller}, {Monkewitz}, {Murphy}, {Ngeow},
  {Nordin}, {Nugent}, {Ofek}, {Patterson}, {Penprase}, {Porter}, {Rauch},
  {Rebbapragada}, {Reiley}, {Rigault}, {Rodriguez}, {van Roestel}, {Rusholme},
  {van Santen}, {Schulze}, {Shupe}, {Singer}, {Soumagnac}, {Stein}, {Surace},
  {Sollerman}, {Szkody}, {Taddia}, {Terek}, {Van Sistine}, {van Velzen},
  {Vestrand}, {Walters}, {Ward}, {Ye}, {Yu}, {Yan}, \&
  {Zolkower}}]{2019PASP..131a8002B}
{Bellm}, E.~C., {Kulkarni}, S.~R., {Graham}, M.~J., {et~al.} 2019, \pasp, 131,
  018002, \dodoi{10.1088/1538-3873/aaecbe}

\bibitem[{{Boch}(2019)}]{2019ASPC..521..487B}
{Boch}, T. 2019, in Astronomical Society of the Pacific Conference Series, Vol.
  521, Astronomical Data Analysis Software and Systems XXVI, ed. M.~{Molinaro},
  K.~{Shortridge}, \& F.~{Pasian}, 487

\bibitem[{{Boch} \& {Fernique}(2014)}]{2014ASPC..485..277B}
{Boch}, T., \& {Fernique}, P. 2014, in Astronomical Society of the Pacific
  Conference Series, Vol. 485, Astronomical Data Analysis Software and Systems
  XXIII, ed. N.~{Manset} \& P.~{Forshay}, 277

\bibitem[{{Bonnarel} {et~al.}(2000){Bonnarel}, {Fernique}, {Bienaym{\'e}},
  {Egret}, {Genova}, {Louys}, {Ochsenbein}, {Wenger}, \&
  {Bartlett}}]{2000A&AS..143...33B}
{Bonnarel}, F., {Fernique}, P., {Bienaym{\'e}}, O., {et~al.} 2000, \aaps, 143,
  33, \dodoi{10.1051/aas:2000331}

\bibitem[{{Brown} {et~al.}(2013){Brown}, {Baliber}, {Bianco}, {Bowman},
  {Burleson}, {Conway}, {Crellin}, {Depagne}, {De Vera}, {Dilday}, {Dragomir},
  {Dubberley}, {Eastman}, {Elphick}, {Falarski}, {Foale}, {Ford}, {Fulton},
  {Garza}, {Gomez}, {Graham}, {Greene}, {Haldeman}, {Hawkins}, {Haworth},
  {Haynes}, {Hidas}, {Hjelstrom}, {Howell}, {Hygelund}, {Lister}, {Lobdill},
  {Martinez}, {Mullins}, {Norbury}, {Parrent}, {Paulson}, {Petry}, {Pickles},
  {Posner}, {Rosing}, {Ross}, {Sand}, {Saunders}, {Shobbrook}, {Shporer},
  {Street}, {Thomas}, {Tsapras}, {Tufts}, {Valenti}, {Vander Horst}, {Walker},
  {White}, \& {Willis}}]{2013PASP..125.1031B}
{Brown}, T.~M., {Baliber}, N., {Bianco}, F.~B., {et~al.} 2013, \pasp, 125,
  1031, \dodoi{10.1086/673168}

\bibitem[{{Calabretta} \& {Greisen}(2002)}]{2002A&A...395.1077C}
{Calabretta}, M.~R., \& {Greisen}, E.~W. 2002, \aap, 395, 1077,
  \dodoi{10.1051/0004-6361:20021327}

\bibitem[{Cannon(2021)}]{ligo-segments}
Cannon, K. 2021, ligo-segments.
\newblock \url{https://git.ligo.org/lscsoft/ligo-segments}

\bibitem[{{Chilingarian} {et~al.}(2004){Chilingarian}, {Bartunov}, {Richter},
  \& {Sigaev}}]{2004ASPC..314..225C}
{Chilingarian}, I., {Bartunov}, O., {Richter}, J., \& {Sigaev}, T. 2004, in
  Astronomical Society of the Pacific Conference Series, Vol. 314, Astronomical
  Data Analysis Software and Systems (ADASS) XIII, ed. F.~{Ochsenbein}, M.~G.
  {Allen}, \& D.~{Egret}, 225

\bibitem[{{Connaughton} {et~al.}(2015){Connaughton}, {Briggs}, {Goldstein},
  {Meegan}, {Paciesas}, {Preece}, {Wilson-Hodge}, {Gibby}, {Greiner}, {Gruber},
  {Jenke}, {Kippen}, {Pelassa}, {Xiong}, {Yu}, {Bhat}, {Burgess}, {Byrne},
  {Fitzpatrick}, {Foley}, {Giles}, {Guiriec}, {van der Horst}, {von Kienlin},
  {McBreen}, {McGlynn}, {Tierney}, \& {Zhang}}]{2015ApJS..216...32C}
{Connaughton}, V., {Briggs}, M.~S., {Goldstein}, A., {et~al.} 2015, \apjs, 216,
  32, \dodoi{10.1088/0067-0049/216/2/32}

\bibitem[{{Coughlin}(2020)}]{2020NatAs...4..550C}
{Coughlin}, M.~W. 2020, Nature Astronomy, 4, 550,
  \dodoi{10.1038/s41550-020-1130-3}

\bibitem[{Danial(2021)}]{adanial_cloc}
Danial, A. 2021, cloc: v1.92, v1.92,  Zenodo, \dodoi{10.5281/zenodo.5760077}

\bibitem[{{Dekany} {et~al.}(2020){Dekany}, {Smith}, {Riddle}, {Feeney},
  {Porter}, {Hale}, {Zolkower}, {Belicki}, {Kaye}, {Henning}, {Walters},
  {Cromer}, {Delacroix}, {Rodriguez}, {Reiley}, {Mao}, {Hover}, {Murphy},
  {Burruss}, {Baker}, {Kowalski}, {Reif}, {Mueller}, {Bellm}, {Graham}, \&
  {Kulkarni}}]{2020PASP..132c8001D}
{Dekany}, R., {Smith}, R.~M., {Riddle}, R., {et~al.} 2020, \pasp, 132, 038001,
  \dodoi{10.1088/1538-3873/ab4ca2}

\bibitem[{Erdogan(2021)}]{postgresql-azure-14-shipped}
Erdogan, O. 2021, How We Shipped PostgreSQL 14 on Azure Within One Day of its
  Release,  Microsoft.
\newblock
  \url{https://techcommunity.microsoft.com/t5/azure-database-for-postgresql/how-we-shipped-postgresql-14-on-azure-within-one-day-of-its/ba-p/2801300}

\bibitem[{{Fernique} {et~al.}(2014){Fernique}, {Boch}, {Donaldson}, {Durand},
  {O'Mullane}, {Reinecke}, \& {Taylor}}]{2014ivoa.spec.0602F}
{Fernique}, P., {Boch}, T., {Donaldson}, T., {et~al.} 2014, {MOC - HEALPix
  Multi-Order Coverage map Version 1.0}, IVOA Recommendation 02 June 2014,
  \dodoi{10.5479/ADS/bib/2014ivoa.spec.0602F}

\bibitem[{{Fernique} {et~al.}(2019){Fernique}, {Boch}, {Donaldson}, {Durand},
  {O'Mullane}, {Reinecke}, \& {Taylor}}]{2019ivoa.spec.1007F}
---. 2019, {MOC - HEALPix Multi-Order Coverage map Version 1.1}, IVOA
  Recommendation 07 October 2019

\bibitem[{{Fernique} {et~al.}(2015){Fernique}, {Allen}, {Boch}, {Oberto},
  {Pineau}, {Durand}, {Bot}, {Cambr{\'e}sy}, {Derriere}, {Genova}, \&
  {Bonnarel}}]{2015A&A...578A.114F}
{Fernique}, P., {Allen}, M.~G., {Boch}, T., {et~al.} 2015, \aap, 578, A114,
  \dodoi{10.1051/0004-6361/201526075}

\bibitem[{{Fernique} {et~al.}(2017){Fernique}, {Allen}, {Boch}, {Donaldson},
  {Durand}, {Ebisawa}, {Michel}, {Salgado}, \& {Stoehr}}]{2017ivoa.spec.0519F}
{Fernique}, P., {Allen}, M., {Boch}, T., {et~al.} 2017, {HiPS - Hierarchical
  Progressive Survey Version 1.0}, IVOA Recommendation 19 May 2017,
  \dodoi{10.5479/ADS/bib/2017ivoa.spec.0519F}

\bibitem[{{F{\"o}rster} {et~al.}(2021){F{\"o}rster}, {Cabrera-Vives},
  {Castillo-Navarrete}, {Est{\'e}vez}, {S{\'a}nchez-S{\'a}ez}, {Arredondo},
  {Bauer}, {Carrasco-Davis}, {Catelan}, {Elorrieta}, {Eyheramendy}, {Huijse},
  {Pignata}, {Reyes}, {Reyes}, {Rodr{\'\i}guez-Mancini}, {Ruz-Mieres},
  {Valenzuela}, {{\'A}lvarez-Maldonado}, {Astorga}, {Borissova}, {Clocchiatti},
  {De Cicco}, {Donoso-Oliva}, {Hern{\'a}ndez-Garc{\'\i}a}, {Graham},
  {Jord{\'a}n}, {Kurtev}, {Mahabal}, {Maureira}, {Mu{\~n}oz-Arancibia},
  {Molina-Ferreiro}, {Moya}, {Palma}, {P{\'e}rez-Carrasco}, {Protopapas},
  {Romero}, {Sabatini-Gacitua}, {S{\'a}nchez}, {San Mart{\'\i}n},
  {Sep{\'u}lveda-Cobo}, {Vera}, \& {Vergara}}]{2021AJ....161..242F}
{F{\"o}rster}, F., {Cabrera-Vives}, G., {Castillo-Navarrete}, E., {et~al.}
  2021, \aj, 161, 242, \dodoi{10.3847/1538-3881/abe9bc}

\bibitem[{{Ginsburg} {et~al.}(2019){Ginsburg}, {Sip{\H{o}}cz}, {Brasseur},
  {Cowperthwaite}, {Craig}, {Deil}, {Guillochon}, {Guzman}, {Liedtke}, {Lian
  Lim}, {Lockhart}, {Mommert}, {Morris}, {Norman}, {Parikh}, {Persson},
  {Robitaille}, {Segovia}, {Singer}, {Tollerud}, {de Val-Borro}, {Valtchanov},
  {Woillez}, {Astroquery Collaboration}, \& {a subset of astropy
  Collaboration}}]{2019AJ....157...98G}
{Ginsburg}, A., {Sip{\H{o}}cz}, B.~M., {Brasseur}, C.~E., {et~al.} 2019, \aj,
  157, 98, \dodoi{10.3847/1538-3881/aafc33}

\bibitem[{{Goldstein} {et~al.}(2020){Goldstein}, {Fletcher}, {Veres}, {Briggs},
  {Cleveland}, {Gibby}, {Hui}, {Bissaldi}, {Burns}, {Hamburg}, {Kienlin},
  {Kocevski}, {Mailyan}, {Malacaria}, {Paciesas}, {Roberts}, \&
  {Wilson-Hodge}}]{2020ApJ...895...40G}
{Goldstein}, A., {Fletcher}, C., {Veres}, P., {et~al.} 2020, \apj, 895, 40,
  \dodoi{10.3847/1538-4357/ab8bdb}

\bibitem[{Google(2022)}]{google-cloud-postgresql-versions}
Google. 2022, Database versions and version policies.
\newblock
  \url{https://web.archive.org/web/20220119055010/https://cloud.google.com/sql/docs/postgres/db-versions}

\bibitem[{{G{\'o}rski} {et~al.}(2005){G{\'o}rski}, {Hivon}, {Banday},
  {Wandelt}, {Hansen}, {Reinecke}, \& {Bartelmann}}]{2005ApJ...622..759G}
{G{\'o}rski}, K.~M., {Hivon}, E., {Banday}, A.~J., {et~al.} 2005, \apj, 622,
  759, \dodoi{10.1086/427976}

\bibitem[{{Graham} {et~al.}(2019){Graham}, {Kulkarni}, {Bellm}, {Adams},
  {Barbarino}, {Blagorodnova}, {Bodewits}, {Bolin}, {Brady}, {Cenko}, {Chang},
  {Coughlin}, {De}, {Eadie}, {Farnham}, {Feindt}, {Franckowiak}, {Fremling},
  {Gezari}, {Ghosh}, {Goldstein}, {Golkhou}, {Goobar}, {Ho}, {Huppenkothen},
  {Ivezi{\'c}}, {Jones}, {Juric}, {Kaplan}, {Kasliwal}, {Kelley}, {Kupfer},
  {Lee}, {Lin}, {Lunnan}, {Mahabal}, {Miller}, {Ngeow}, {Nugent}, {Ofek},
  {Prince}, {Rauch}, {van Roestel}, {Schulze}, {Singer}, {Sollerman}, {Taddia},
  {Yan}, {Ye}, {Yu}, {Barlow}, {Bauer}, {Beck}, {Belicki}, {Biswas}, {Brinnel},
  {Brooke}, {Bue}, {Bulla}, {Burruss}, {Connolly}, {Cromer}, {Cunningham},
  {Dekany}, {Delacroix}, {Desai}, {Duev}, {Feeney}, {Flynn}, {Frederick},
  {Gal-Yam}, {Giomi}, {Groom}, {Hacopians}, {Hale}, {Helou}, {Henning},
  {Hover}, {Hillenbrand}, {Howell}, {Hung}, {Imel}, {Ip}, {Jackson}, {Kaspi},
  {Kaye}, {Kowalski}, {Kramer}, {Kuhn}, {Landry}, {Laher}, {Mao}, {Masci},
  {Monkewitz}, {Murphy}, {Nordin}, {Patterson}, {Penprase}, {Porter},
  {Rebbapragada}, {Reiley}, {Riddle}, {Rigault}, {Rodriguez}, {Rusholme}, {van
  Santen}, {Shupe}, {Smith}, {Soumagnac}, {Stein}, {Surace}, {Szkody}, {Terek},
  {Van Sistine}, {van Velzen}, {Vestrand}, {Walters}, {Ward}, {Zhang}, \&
  {Zolkower}}]{2019PASP..131g8001G}
{Graham}, M.~J., {Kulkarni}, S.~R., {Bellm}, E.~C., {et~al.} 2019, \pasp, 131,
  078001, \dodoi{10.1088/1538-3873/ab006c}

\bibitem[{Greco {et~al.}(2019)Greco, Branchesi, Chassande-Mottin, Coughlin,
  Stratta, Dálya, Hemming, Rei, Brocato, Fernique, Boch, Derriere, Baumann,
  Genova, \& Allen}]{Greco:20191+}
Greco, G., Branchesi, M., Chassande-Mottin, E., {et~al.} 2019, in Proceedings
  of The New Era of Multi-Messenger Astrophysics {\textemdash}
  PoS(Asterics2019), Vol. 357, 031, \dodoi{10.22323/1.357.0031}

\bibitem[{{Greisen} \& {Calabretta}(2002)}]{2002A&A...395.1061G}
{Greisen}, E.~W., \& {Calabretta}, M.~R. 2002, \aap, 395, 1061,
  \dodoi{10.1051/0004-6361:20021326}

\bibitem[{{Harris} {et~al.}(2020){Harris}, {Millman}, {van der Walt},
  {Gommers}, {Virtanen}, {Cournapeau}, {Wieser}, {Taylor}, {Berg}, {Smith},
  {Kern}, {Picus}, {Hoyer}, {van Kerkwijk}, {Brett}, {Haldane}, {del R{\'\i}o},
  {Wiebe}, {Peterson}, {G{\'e}rard-Marchant}, {Sheppard}, {Reddy}, {Weckesser},
  {Abbasi}, {Gohlke}, \& {Oliphant}}]{2020Natur.585..357H}
{Harris}, C.~R., {Millman}, K.~J., {van der Walt}, S.~J., {et~al.} 2020, \nat,
  585, 357, \dodoi{10.1038/s41586-020-2649-2}

\bibitem[{Hellerstein {et~al.}(1995)Hellerstein, Naughton, \& Pfeffer}]{GiST}
Hellerstein, J.~M., Naughton, J.~F., \& Pfeffer, A. 1995, in Proceedings of the
  21th International Conference on Very Large Data Bases, VLDB '95 (San
  Francisco, CA, USA: Morgan Kaufmann Publishers Inc.), 562–573

\bibitem[{{Huchra} {et~al.}(2012){Huchra}, {Macri}, {Masters}, {Jarrett},
  {Berlind}, {Calkins}, {Crook}, {Cutri}, {Erdo{\v{g}}du}, {Falco}, {George},
  {Hutcheson}, {Lahav}, {Mader}, {Mink}, {Martimbeau}, {Schneider},
  {Skrutskie}, {Tokarz}, \& {Westover}}]{2012ApJS..199...26H}
{Huchra}, J.~P., {Macri}, L.~M., {Masters}, K.~L., {et~al.} 2012, \apjs, 199,
  26, \dodoi{10.1088/0067-0049/199/2/26}

\bibitem[{{Hunter}(2007)}]{2007CSE.....9...90H}
{Hunter}, J.~D. 2007, Computing in Science and Engineering, 9, 90,
  \dodoi{10.1109/MCSE.2007.55}

\bibitem[{{Ivezi{\'c}} {et~al.}(2019){Ivezi{\'c}}, {Kahn}, {Tyson}, {Abel},
  {Acosta}, {Allsman}, {Alonso}, {AlSayyad}, {Anderson}, {Andrew}, \&
  et~al.}]{2019ApJ...873..111I}
{Ivezi{\'c}}, {\v{Z}}., {Kahn}, S.~M., {Tyson}, J.~A., {et~al.} 2019, \apj,
  873, 111, \dodoi{10.3847/1538-4357/ab042c}

\bibitem[{{Kasliwal} {et~al.}(2019){Kasliwal}, {Cannella}, {Bagdasaryan},
  {Hung}, {Feindt}, {Singer}, {Coughlin}, {Fremling}, {Walters}, {Duev},
  {Itoh}, \& {Quimby}}]{2019PASP..131c8003K}
{Kasliwal}, M.~M., {Cannella}, C., {Bagdasaryan}, A., {et~al.} 2019, \pasp,
  131, 038003, \dodoi{10.1088/1538-3873/aafbc2}

\bibitem[{{Kasliwal} {et~al.}(2020){Kasliwal}, {Anand}, {Ahumada}, {Stein},
  {Carracedo}, {Andreoni}, {Coughlin}, {Singer}, {Kool}, {De}, {Kumar},
  {AlMualla}, {Yao}, {Bulla}, {Dobie}, {Reusch}, {Perley}, {Cenko}, {Bhalerao},
  {Kaplan}, {Sollerman}, {Goobar}, {Copperwheat}, {Bellm}, {Anupama}, {Corsi},
  {Nissanke}, {Agudo}, {Bagdasaryan}, {Barway}, {Belicki}, {Bloom}, {Bolin},
  {Buckley}, {Burdge}, {Burruss}, {Caballero-Garc{\'\i}a}, {Cannella},
  {Castro-Tirado}, {Cook}, {Cooke}, {Cunningham}, {Dahiwale}, {Deshmukh},
  {Dichiara}, {Duev}, {Dutta}, {Feeney}, {Franckowiak}, {Frederick},
  {Fremling}, {Gal-Yam}, {Gatkine}, {Ghosh}, {Goldstein}, {Golkhou}, {Graham},
  {Graham}, {Hankins}, {Helou}, {Hu}, {Ip}, {Jaodand}, {Karambelkar}, {Kong},
  {Kowalski}, {Khandagale}, {Kulkarni}, {Kumar}, {Laher}, {Li}, {Mahabal},
  {Masci}, {Miller}, {Mogotsi}, {Mohite}, {Mooley}, {Mroz}, {Newman}, {Ngeow},
  {Oates}, {Patil}, {Pandey}, {Pavana}, {Pian}, {Riddle},
  {S{\'a}nchez-Ram{\'\i}rez}, {Sharma}, {Singh}, {Smith}, {Soumagnac},
  {Taggart}, {Tan}, {Tzanidakis}, {Troja}, {Valeev}, {Walters}, {Waratkar},
  {Webb}, {Yu}, {Zhang}, {Zhou}, \& {Zolkower}}]{2020ApJ...905..145K}
{Kasliwal}, M.~M., {Anand}, S., {Ahumada}, T., {et~al.} 2020, \apj, 905, 145,
  \dodoi{10.3847/1538-4357/abc335}

\bibitem[{Koposov(2020)}]{pghealpix}
Koposov, S. 2020, pg\_healpix.
\newblock \url{https://github.com/segasai/pg_healpix}

\bibitem[{{Koposov} \& {Bartunov}(2006)}]{2006ASPC..351..735K}
{Koposov}, S., \& {Bartunov}, O. 2006, in Astronomical Society of the Pacific
  Conference Series, Vol. 351, Astronomical Data Analysis Software and Systems
  XV, ed. C.~{Gabriel}, C.~{Arviset}, D.~{Ponz}, \& S.~{Enrique}, 735

\bibitem[{{Koposov} \& {Bartunov}(2019)}]{2019ascl.soft05008K}
{Koposov}, S., \& {Bartunov}, O. 2019, {Q3C: A PostgreSQL package for spatial
  queries and cross-matches of large astronomical catalogs}.
\newblock \doeprint{1905.008}

\bibitem[{{Landais} {et~al.}(2013){Landais}, {Ochsenbein}, \&
  {Simon}}]{2013ASPC..475..227L}
{Landais}, G., {Ochsenbein}, F., \& {Simon}, A. 2013, in Astronomical Society
  of the Pacific Conference Series, Vol. 475, Astronomical Data Analysis
  Software and Systems XXII, ed. D.~N. {Friedel}, 227

\bibitem[{{Lang} {et~al.}(2010){Lang}, {Hogg}, {Mierle}, {Blanton}, \&
  {Roweis}}]{2010AJ....139.1782L}
{Lang}, D., {Hogg}, D.~W., {Mierle}, K., {Blanton}, M., \& {Roweis}, S. 2010,
  \aj, 139, 1782, \dodoi{10.1088/0004-6256/139/5/1782}

\bibitem[{{Las Cumbres Observatory}(2019)}]{tom-toolkit-workshop}
{Las Cumbres Observatory}. 2019, in TOM Toolkit Workshop.
\newblock \url{https://lco.global/workshops/tom-toolkit-community-workshop/}

\bibitem[{{LIGO Scientific Collaboration} {et~al.}(2015){LIGO Scientific
  Collaboration}, {Aasi}, {Abbott}, {Abbott}, {Abbott}, {Abernathy}, {Ackley},
  {Adams}, {Adams}, {Addesso}, \& et~al.}]{2015CQGra..32g4001L}
{LIGO Scientific Collaboration}, {Aasi}, J., {Abbott}, B.~P., {et~al.} 2015,
  Classical and Quantum Gravity, 32, 074001,
  \dodoi{10.1088/0264-9381/32/7/074001}

\bibitem[{{Martinez-Castellanos} {et~al.}(2021){Martinez-Castellanos},
  {Singer}, {Burns}, {Tak}, {Joens}, {Racusin}, \&
  {Perkins}}]{2021arXiv211111240M}
{Martinez-Castellanos}, I., {Singer}, L.~P., {Burns}, E., {et~al.} 2021, arXiv
  e-prints, arXiv:2111.11240.
\newblock \doarXiv{2111.11240}

\bibitem[{{Masci} {et~al.}(2019){Masci}, {Laher}, {Rusholme}, {Shupe}, {Groom},
  {Surace}, {Jackson}, {Monkewitz}, {Beck}, {Flynn}, {Terek}, {Landry},
  {Hacopians}, {Desai}, {Howell}, {Brooke}, {Imel}, {Wachter}, {Ye}, {Lin},
  {Cenko}, {Cunningham}, {Rebbapragada}, {Bue}, {Miller}, {Mahabal}, {Bellm},
  {Patterson}, {Juri{\'c}}, {Golkhou}, {Ofek}, {Walters}, {Graham}, {Kasliwal},
  {Dekany}, {Kupfer}, {Burdge}, {Cannella}, {Barlow}, {Van Sistine}, {Giomi},
  {Fremling}, {Blagorodnova}, {Levitan}, {Riddle}, {Smith}, {Helou}, {Prince},
  \& {Kulkarni}}]{2019PASP..131a8003M}
{Masci}, F.~J., {Laher}, R.~R., {Rusholme}, B., {et~al.} 2019, \pasp, 131,
  018003, \dodoi{10.1088/1538-3873/aae8ac}

\bibitem[{{Matheson} {et~al.}(2021){Matheson}, {Stubens}, {Wolf}, {Lee},
  {Narayan}, {Saha}, {Scott}, {Soraisam}, {Bolton}, {Hauger}, {Silva},
  {Kececioglu}, {Scheidegger}, {Snodgrass}, {Aleo}, {Evans-Jacquez}, {Singh},
  {Wang}, {Yang}, \& {Zhao}}]{2021AJ....161..107M}
{Matheson}, T., {Stubens}, C., {Wolf}, N., {et~al.} 2021, \aj, 161, 107,
  \dodoi{10.3847/1538-3881/abd703}

\bibitem[{{Meegan} {et~al.}(2009){Meegan}, {Lichti}, {Bhat}, {Bissaldi},
  {Briggs}, {Connaughton}, {Diehl}, {Fishman}, {Greiner}, {Hoover}, {van der
  Horst}, {von Kienlin}, {Kippen}, {Kouveliotou}, {McBreen}, {Paciesas},
  {Preece}, {Steinle}, {Wallace}, {Wilson}, \&
  {Wilson-Hodge}}]{2009ApJ...702..791M}
{Meegan}, C., {Lichti}, G., {Bhat}, P.~N., {et~al.} 2009, \apj, 702, 791,
  \dodoi{10.1088/0004-637X/702/1/791}

\bibitem[{Mehlhorn \& Sanders(2008)}]{mehlhorn2008algorithms}
Mehlhorn, K., \& Sanders, P. 2008, Algorithms and Data Structures: The Basic
  Toolbox, SpringerLink: Springer e-Books (Springer Berlin Heidelberg).
\newblock \url{https://books.google.com/books?id=H2BDafez-A0C}

\bibitem[{{M{\"o}ller} {et~al.}(2021){M{\"o}ller}, {Peloton}, {Ishida},
  {Arnault}, {Bachelet}, {Blaineau}, {Boutigny}, {Chauhan}, {Gangler},
  {Hernandez}, {Hrivnac}, {Leoni}, {Leroy}, {Moniez}, {Pateyron}, {Ramparison},
  {Turpin}, {Ansari}, {Allam}, {Bajat}, {Biswas}, {Boucaud}, {Bregeon},
  {Campagne}, {Cohen-Tanugi}, {Coleiro}, {Dornic}, {Fouchez}, {Godet}, {Gris},
  {Karpov}, {Nebot Gomez-Moran}, {Neveu}, {Plaszczynski}, {Savchenko}, \&
  {Webb}}]{2021MNRAS.501.3272M}
{M{\"o}ller}, A., {Peloton}, J., {Ishida}, E. E.~O., {et~al.} 2021, \mnras,
  501, 3272, \dodoi{10.1093/mnras/staa3602}

\bibitem[{{Nordin} {et~al.}(2019){Nordin}, {Brinnel}, {van Santen}, {Bulla},
  {Feindt}, {Franckowiak}, {Fremling}, {Gal-Yam}, {Giomi}, {Kowalski},
  {Mahabal}, {Miranda}, {Rauch}, {Reusch}, {Rigault}, {Schulze}, {Sollerman},
  {Stein}, {Yaron}, {van Velzen}, \& {Ward}}]{2019A&A...631A.147N}
{Nordin}, J., {Brinnel}, V., {van Santen}, J., {et~al.} 2019, \aap, 631, A147,
  \dodoi{10.1051/0004-6361/201935634}

\bibitem[{Obe \& Hsu(2021)}]{obe2021postgis}
Obe, R., \& Hsu, L. 2021, PostGIS in Action, Third Edition (Manning).
\newblock \url{https://books.google.com/books?id=6PY8EAAAQBAJ}

\bibitem[{{Pence} {et~al.}(2010){Pence}, {Chiappetti}, {Page}, {Shaw}, \&
  {Stobie}}]{2010A&A...524A..42P}
{Pence}, W.~D., {Chiappetti}, L., {Page}, C.~G., {Shaw}, R.~A., \& {Stobie}, E.
  2010, \aap, 524, A42, \dodoi{10.1051/0004-6361/201015362}

\bibitem[{{Petrov} {et~al.}(2022){Petrov}, {Singer}, {Coughlin}, {Kumar},
  {Almualla}, {Anand}, {Bulla}, {Dietrich}, {Foucart}, \&
  {Guessoum}}]{2022ApJ...924...54P}
{Petrov}, P., {Singer}, L.~P., {Coughlin}, M.~W., {et~al.} 2022, \apj, 924, 54,
  \dodoi{10.3847/1538-4357/ac366d}

\bibitem[{Raen(2021)}]{pitt-google-broker}
Raen, T. 2021, Pitt-Google Alert Broker.
\newblock \url{https://github.com/mwvgroup/Pitt-Google-Broker}

\bibitem[{{Reinecke} \& {Hivon}(2015)}]{2015A&A...580A.132R}
{Reinecke}, M., \& {Hivon}, E. 2015, \aap, 580, A132,
  \dodoi{10.1051/0004-6361/201526549}

\bibitem[{{Robitaille} {et~al.}(2020){Robitaille}, {Deil}, \&
  {Ginsburg}}]{2020ascl.soft11023R}
{Robitaille}, T., {Deil}, C., \& {Ginsburg}, A. 2020, {reproject: Python-based
  astronomical image reprojection}.
\newblock \doeprint{2011.023}

\bibitem[{{Sambruna} {et~al.}(2021){Sambruna}, {Schlieder}, {Kocevski},
  {Caputo}, {Hui}, {Markwardt}, {Powell}, {Racusin}, {Roberts}, {Singer},
  {Smale}, {Venters}, \& {Wilson-Hodge}}]{2021arXiv210910841S}
{Sambruna}, R.~M., {Schlieder}, J.~E., {Kocevski}, D., {et~al.} 2021, arXiv
  e-prints, arXiv:2109.10841.
\newblock \doarXiv{2109.10841}

\bibitem[{Singer {et~al.}(2021)Singer, Goldstein, Crellin-Quick, Parazin, \&
  Coughlin}]{healpix-alchemy-1.0.1}
Singer, L., Goldstein, D., Crellin-Quick, A., Parazin, B., \& Coughlin, M.
  2021, skyportal/healpix-alchemy: Version 1.0.1, v1.0.1,  Zenodo,
  \dodoi{10.5281/zenodo.5768564}

\bibitem[{{Singer} \& {Price}(2016)}]{2016PhRvD..93b4013S}
{Singer}, L.~P., \& {Price}, L.~R. 2016, \prd, 93, 024013,
  \dodoi{10.1103/PhysRevD.93.024013}

\bibitem[{{Singer} {et~al.}(2016){Singer}, {Chen}, {Holz}, {Farr}, {Price},
  {Raymond}, {Cenko}, {Gehrels}, {Cannizzo}, {Kasliwal}, {Nissanke},
  {Coughlin}, {Farr}, {Urban}, {Vitale}, {Veitch}, {Graff}, {Berry},
  {Mohapatra}, \& {Mandel}}]{2016ApJ...829L..15S}
{Singer}, L.~P., {Chen}, H.-Y., {Holz}, D.~E., {et~al.} 2016, \apjl, 829, L15,
  \dodoi{10.3847/2041-8205/829/1/L15}

\bibitem[{{Smith} {et~al.}(2019){Smith}, {Williams}, {Young}, {Ibsen},
  {Smartt}, {Lawrence}, {Morris}, {Voutsinas}, \&
  {Nicholl}}]{2019RNAAS...3...26S}
{Smith}, K.~W., {Williams}, R.~D., {Young}, D.~R., {et~al.} 2019, Research
  Notes of the American Astronomical Society, 3, 26,
  \dodoi{10.3847/2515-5172/ab020f}

\bibitem[{Stonebraker \& Rowe(1986)}]{PostgreSQL}
Stonebraker, M., \& Rowe, L.~A. 1986, in Proceedings of the 1986 ACM SIGMOD
  International Conference on Management of Data, SIGMOD '86 (New York, NY,
  USA: Association for Computing Machinery), 340–355,
  \dodoi{10.1145/16894.16888}

\bibitem[{Stovner \& Sætrom(2019)}]{10.1093/bioinformatics/btz615}
Stovner, E.~B., \& Sætrom, P. 2019, Bioinformatics, 36, 918,
  \dodoi{10.1093/bioinformatics/btz615}

\bibitem[{{Szalay} {et~al.}(2007){Szalay}, {Gray}, {Fekete}, {Kunszt}, {Kukol},
  \& {Thakar}}]{2007cs........1164S}
{Szalay}, A.~S., {Gray}, J., {Fekete}, G., {et~al.} 2007, arXiv e-prints,
  cs/0701164.
\newblock \doarXiv{cs/0701164}

\bibitem[{{van der Walt} {et~al.}(2019){van der Walt}, {Crellin-Quick}, \&
  {Bloom}}]{2019JOSS....4.1247V}
{van der Walt}, S., {Crellin-Quick}, A., \& {Bloom}, J. 2019, The Journal of
  Open Source Software, 4, 1247, \dodoi{10.21105/joss.01247}

\bibitem[{{Wyatt} {et~al.}(2020){Wyatt}, {Tohuvavohu}, {Arcavi}, {Lundquist},
  {Howell}, \& {Sand}}]{2020ApJ...894..127W}
{Wyatt}, S.~D., {Tohuvavohu}, A., {Arcavi}, I., {et~al.} 2020, \apj, 894, 127,
  \dodoi{10.3847/1538-4357/ab855e}

\bibitem[{{Zonca} {et~al.}(2019){Zonca}, {Singer}, {Lenz}, {Reinecke},
  {Rosset}, {Hivon}, \& {Gorski}}]{2019JOSS....4.1298Z}
{Zonca}, A., {Singer}, L., {Lenz}, D., {et~al.} 2019, The Journal of Open
  Source Software, 4, 1298, \dodoi{10.21105/joss.01298}

\end{thebibliography}
\bibliographystyle{aasjournal}

\end{document}